\documentclass[conference]{IEEEtran}
\IEEEoverridecommandlockouts
\usepackage[english]{babel}
\usepackage[utf8x]{inputenc}
\usepackage{amsmath}
\usepackage{amssymb}
\usepackage{float}
\usepackage{graphicx}
\usepackage{array}
\usepackage{multirow}
\usepackage{longtable}
\usepackage{hyperref}
\usepackage{authblk}
\usepackage{adjustbox}
\usepackage{pgfplots}
\usepackage{booktabs}
\usepackage[merge,square,sort&compress,numbers]{natbib}
\pgfplotsset{compat=1.12}
\usepackage{tabularx}
    \newcolumntype{L}{>{\raggedright\arraybackslash}X}

\usepackage{pifont}%
\usepackage{url}

\usepackage[flushleft]{threeparttable}
\usepackage[inline,ignoremode]{trackchanges} % Documentation of this package can be found at http://trackchanges.sourceforge.net/
\addeditor{A. Pasdar} % For each editor, repeat this command with their initials. Monroe is added by default.
\newcommand{\xmark}{\ding{55}}%

\title{Blockchain Oracle Design Patterns}
\author{Amirmohammad Pasdar$^{1,2}$, Zhongli Dong$^{1,3}$, Young Choon Lee$^{2}$\\
\thanks{$^{1}$ Amirmohammad Pasdar (amir@aglive.com) and Zhongli Dong (andrew@aglive.com) are with Aglive Lab, 8 Hadenfeld Ave, Macquarie Park NSW 2113, Australia.}
\thanks{$^{2}$ Amirmohammad Pasdar (amirmohammad.pasdar@hdr.mq.edu.au) and Young Choon Lee (young.lee@mq.edu.au) are with Department of Computing, Macquarie University, Sydney NSW 2109, Australia.}
\thanks{$^{3}$ Zhongli Dong (zhongli.dong@sydney.edu.au) is with Center for Distributed and High Performance Computing, School of Computer Science, The University of Sydney, NSW 2006, Australia.}

}

\date{} % You can write in any date within the brackets.
\usepackage{perpage} %the perpage package
\MakePerPage{footnote} %the perpage package command
\begin{document}
\sloppy
\maketitle

\begin{abstract}
Blockchain is a form of distributed ledger technology (DLT) where data is shared among users connected over the internet. Transactions are data state changes on the blockchain that are permanently recorded in a secure and transparent way without the need of a third party. Besides, the introduction of smart contracts to the blockchain has added programmability to the blockchain and revolutionized the software ecosystem leading toward decentralized applications (DApps) attracting businesses and organizations to employ this technology. Although promising, blockchains and smart contracts have no access to the external systems (i.e., off-chain) where real-world data and events resides; consequently, the usability of smart contracts in terms of performance and programmability would be limited to the on-chain data. Hence, \emph{blockchain oracles} are introduced to mitigate the issue and are defined as trusted third-party services that send and verify the external information (i.e., feedback) and submit it to smart contracts for triggering state changes in the blockchain. In this paper, we will study and analyze blockchain oracles with regard to how they provide feedback to the blockchain and smart contracts. We classify the blockchain oracle techniques into two major groups such as voting-based strategies and reputation-based ones. The former mainly relies on participants' stakes for outcome finalization while the latter considers reputation in conjunction with authenticity proof mechanisms for data correctness and integrity. We then provide a structured description of patterns in detail for each classification and discuss research directions in the end.
\end{abstract}


\section{Introduction}

A blockchain is a form of distributed ledger technology where transactions are duplicated and saved onto a large number of nodes. Transactions are defined as a set of data packages for storing monetary value, parameters, and function call results, and their integrity is ensured by cryptographic techniques. They are collected in the form of blocks where there are immutable records, and each block is linked to the succeeding block through the hash. Blocks are appended to the ledger by a means of consensus algorithms such as Proof of Work (PoW), Proof of Stake (PoS), or Practical Byzantine Fault Tolerance (PBFT) \cite{orc36}. Blockchain technology is a good fit for decentralized finance (DeFi) or dealing with data integrity, e.g., supplying products \cite{orc29} or food security \cite{foodbc}. 

Smart contracts are programs (i.e., functions and states) residing at a specific address executed on the blockchain to digitally facilitate the transaction process. They are event-driven, self-executable, resistant to tamper, and they can consume transaction fees based on the complexity of code (e.g., gas for deployment in the Ethereum blockchain), and only use resources available on the blockchain network \cite{orc18}. Smart contracts are encoded and compiled into bytecode and upon deployment, they are given unique addresses and saved across all connected nodes in the network. 

\begin{figure}
    \centering
    \includegraphics[keepaspectratio,scale=0.16]{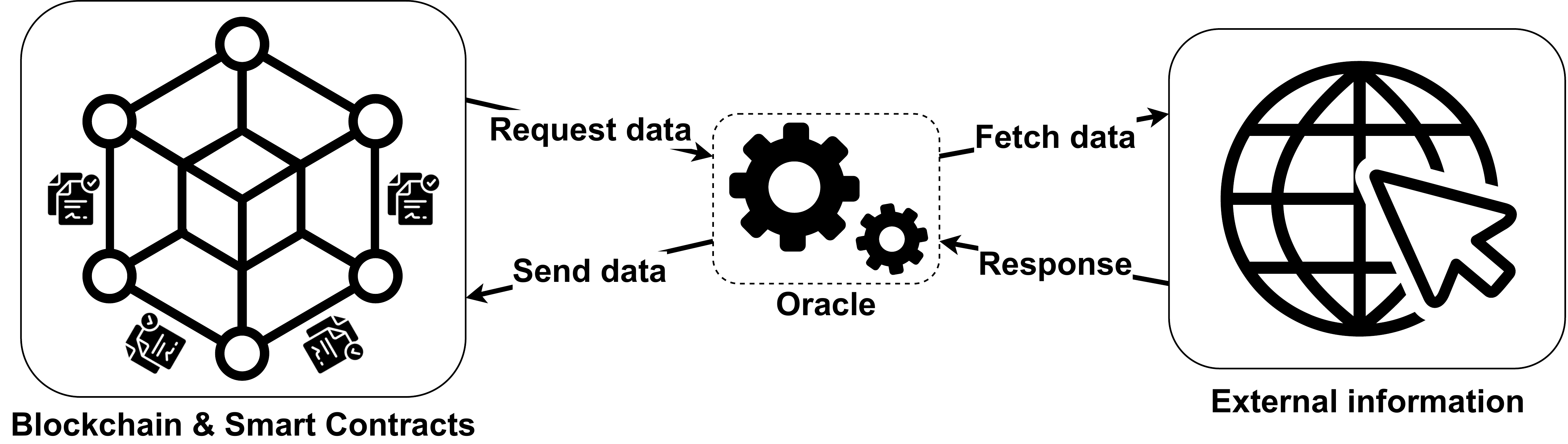}
    \caption{The oracle role as trusted third-party services  for sending and verifying the external information. Blockchain oracles can be single or a pool of oracles interacting with the external world.}
    \label{fig:oracle}
\end{figure}

Smart contracts and blockchain \emph{do not} have access to the information outside of their networks (i.e., off-chain data). The blockchain in fact is an enclosed system where interactions are limited to the data available on it. Hence, it is still an open practical problem referred to as the ``oracle problem'' that is defined as how real-world data can be transferred into/from the blockchain. If the blockchain is assumed as a component of a larger software system \cite{orc22}, smart contracts need the external information relevant to contractual agreements or practical applications, e.g., data availability verification for decentralized applications or adjudication mechanisms. Oracles (also known as \emph{data feeds}) shown in Figure \ref{fig:oracle} act as trusted third-party services that send and verify the external information and submit it to smart contracts to trigger state changes in the blockchain. Oracles may not only relay information to the smart contracts, but also send it to external resources. They are simply contracts on the blockchain for serving data requests by other contracts \cite{orc9}. Without oracles, smart contracts would have limited connectability; hence, they are vital for the blockchain ecosystem due to broadening the scope of smart contracts operation. 

There have been several surveys on blockchain oracles \cite{orc21,orc23, orc31,orc32,orc33,orc35} each of which studied particular aspect(s) of oracles. Muhlberger et al., \cite{orc21} examine oracles from two dimensions which are data flow direction and the data initiator. % of the data. Based on Muhlberger's study, the data flow is defined as pull/push data onto/from the blockchain which depends on where the data is initiated; on-chain or off-chain. data annotation for machine learning,
Heiss et al., \cite{orc21} present a key requirement set for trustworthy data on-chain, and how related challenges should be addressed. Beniiche \cite{orc31} describes widely used blockchain oracles and human oracles, and Al-Breiki et al., \cite{orc32} study the trust used in the leading blockchain oracles. Xu et al., \cite{orc33} discuss oracle roles and provide the benefits and drawbacks, and Mammadzada et al., \cite{orc35} provide general characteristics of a blockchain framework to be considered when designing blockchain-based applications.

This study reviews and analyzes blockchain oracles from a technical perspective that previous studies have overlooked. We analyze the blockchain oracles in terms of providing final feedback (i.e., outcome) to the smart contracts, and we mainly seek an answer for the following research question (RQ);

\textit{RQ: How are blockchain oracles designed to provide the outcome to smart contracts?}

RQ fundamentally contributes to providing deep insights into the design, development requirements, and usage of blockchain oracles. Hence, to properly answer the RQ1, we conduct this research through the Multivocal Literature Review (MLR) technique defined as a form of Systematic Literature Review (SLM) in which ``grey'' literature and ``white papers'' are included \cite{mlr}. The initial search keywords are selected based on a combination of \emph{blockchain oracles}, \emph{data feed}, and/or \emph{smart contracts}, \emph{design} and/or \emph{pattern} for covering the vast majority of related studies. During the paper collection, snowballing technique \cite{snowballing} is employed to collect relevant studies for the literature review. Several digital libraries are used to extract a fine set of papers on the proposed topic such as IEEE Xplore\footnote{https://ieeexplore.ieee.org/Xplore/home.jsp}, ScienceDirect\footnote{https://www.sciencedirect.com/}, ACM Digital Library\footnote{https://dl.acm.org}, DBLP:Computer Science Bibliography\footnote{https://dblp.org/}, and Google Scholar\footnote{https://scholar.google.com.au/}.  

\begin{figure}
    \centering
    \includegraphics[keepaspectratio,scale=0.42]{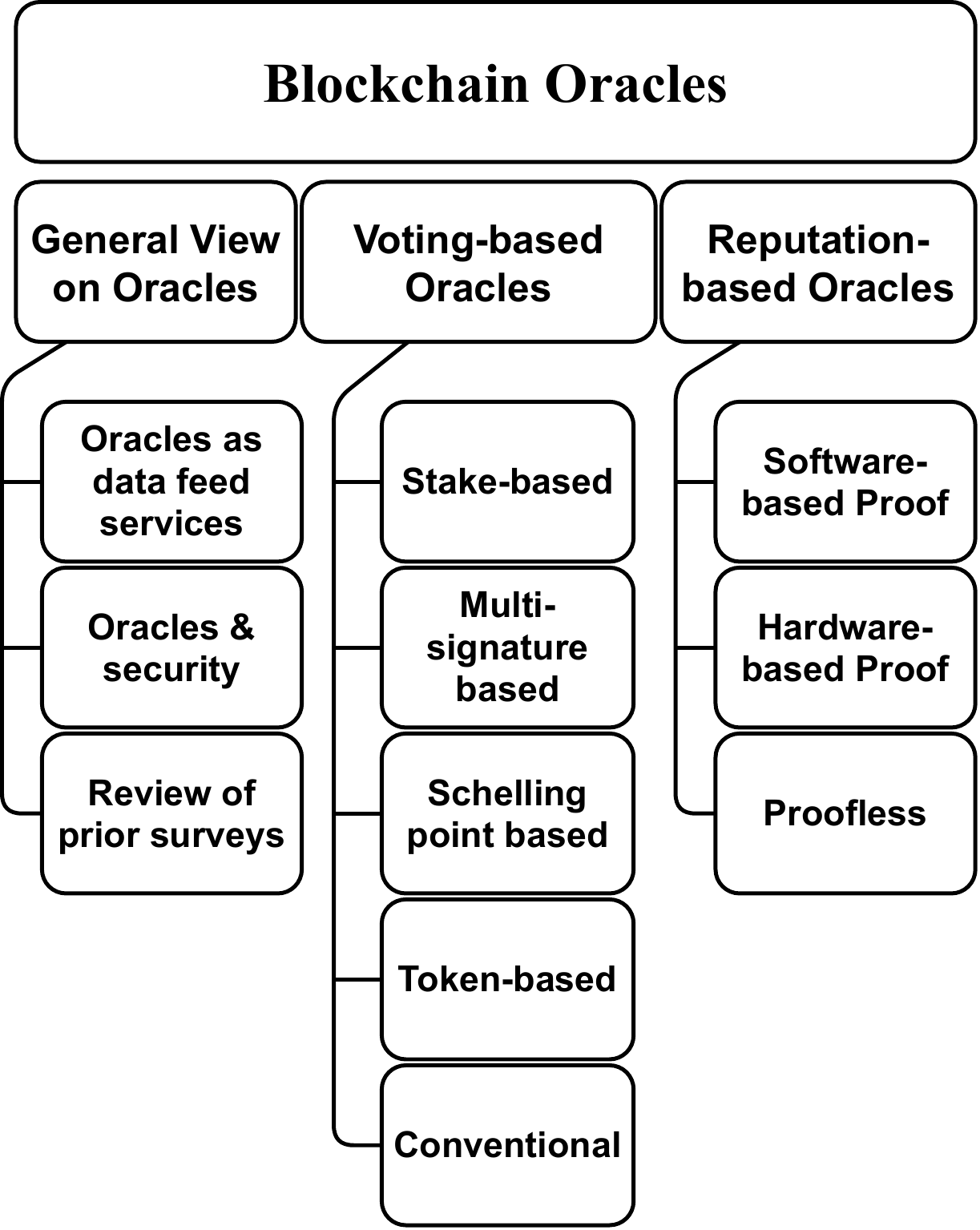}
    \caption{The structure of research study.}
    \label{fig:paper_struc}
\end{figure}

Finally, with the ever-increasing usage of blockchain, the need for real-world data for extending the blockchain usability while mitigating the potential risk of data manipulation is exponentially increasing \cite{iota}. Hence, this study intends to fill the gap by presenting techniques, challenges, advantages and disadvantages of existing blockchain oracle design patterns through a comprehensive review of cutting-edge studies. In this regard, we categorize the collected research studies on either the oracle design or oracle usage into two major groups with respect to the \emph{monetary incentives} employed in the blockchain oracle design. Incentive mechanisms are in place to encourage users to participate in the network governance and improve the blockchain security and functionality. The first group is oracles that employ \emph{voting-based} strategies for data aggregation and outcome. The second group leverages \emph{reputation-based} strategy to select the oracle for reporting back the outcome to the requester. They may employ authenticity proof mechanisms to prove the integrity and correctness of obtained data from external resources. Figure \ref{fig:paper_struc} shows the structure of this research survey.

The paper is organized as follows: Section \ref{sec:blockchain} briefly overviews the survey papers on blockchain oracles. In Section \ref{sec:voting_oracle} we present voting-based oracles as a way of providing oracle feedback which is followed by Section \ref{sec:reputaion} in which authenticity proof mechanisms are also studied. Section \ref{sec:frd} presents future research directions, and we conclude the paper in Section \ref{sec:conclusion}.

% \section{Blockchain Oracles}
\section{General View on Oracles \& Prior Research}
\label{sec:blockchain}
This section presents a general classification of oracles as data feed services, issues, and discussions on prior survey research on blockchain oracles. 

\subsection{Oracles as Data Feed Services}

Oracles (or data feed services) respond to queries which can be in the form of ``What is the current ETH\footnote{Ether (ETH) is the native token of the Ethereum blockchain.}/USD exchange rate?'', and oracles may consult with different sources (or a single source) to fetch the required information and return it to the smart contracts. Oracles can also be \emph{computation oracles} such as Truebit \cite{truebit} or Provable \cite{orc8} that perform user-defined computation-intensive tasks off-chain, providing computing power to the blockchains, and enabling decentralized token economy. In addition, a general view of blockchain oracles facilitates its classification which can be fallen into four major groups: (1) \emph{source}; the origin of data, (2) \emph{information direction}; inbound or outbound, (3) \emph{trust}; centralized or decentralized, and (4) \emph{design pattern} \cite{orc32, orc31}.  

\emph{Oracle sources} can be chosen from (a) software oracles where data comes from online sources (e.g., online servers), (b) hardware oracles where data comes from the physical world (e.g., sensors), and (c) human oracles that are also responsible for verifying the authenticity of information and its translation into smart contracts. \emph{Information direction} means the way information flows; from/to external resources, e.g., a smart lock is an outbound oracle that once the payment is received at the address, unlocks the lock in the physical world. There is the concept of \emph{trust} that can be centralized or decentralized. Centralized oracles are efficient but they can be risky because a single entity provides information, controls the oracles, and a failure makes the contracts less resilient to vulnerabilities and attacks. In contrast, decentralized oracles (i.e., consensus-based oracles) increase the reliability of the information provided to the smart contracts by querying multiple resources. It should be noted that an oracle is considered decentralized if it is permissionless- users can join or leave, and every user has an equal privilege \cite{orc25}. Finally, \emph{design patterns} are defined as (a) request-response when the data space is huge and can be implemented as on-chain smart contracts and initiated on-chain, or off-chain oracles for monitoring, retrieving, and returning data (b) publish-subscribe when the data is expected to change, e.g., RSS feeds, and (c) immediate read when the data is required for an immediate decision.

\subsection{Oracles \& Security}

Data feed services to the smart contracts may have desired properties such as how easily they are parsable, adopted, and deployed as well as their authenticity, and being non-equivocation (i.e., being unable to modify or delete data when becomes published) \cite{pdfs}. However, it may come with issues that emerge from putting \emph{trust} in a single third party which is represented as a single \emph{point of failure} because external malicious actors can break into a single system and alter or delete facts. Oracles are prone to be hacked; their process is vague, they can be bribed, and may not be stable \cite{orc12}. Also, smart contracts lack \emph{direct} network access, and the use of transport layer security (TLS) to fetch information while keeping data untampered during transmission is not enough \cite{tlsn}. Hence, mechanisms should exist to digitally sign the data for verification, and in this regard, oracles are neither tamper-resistant nor trustless. 

Oracles do not mainly hold security properties of native blockchain protocols, however, the correctness of the data can be attested through authenticity proof mechanisms, e.g., software-based \cite{orc20,tlsn} or hardware-based approaches \cite{ledgerproof,orc9}. Although an ideal oracle is hard to achieve, oracles must provide the same level of security in proportion to the blockchain they support in the form of integrity, confidentiality, and availability. High economic security that is defined as financial resources required for compromising a network, should be in a way that compromising a network would not be beneficial if the financial benefits are not higher than the cost. Hence, the higher the decentralized oracle platform degree, the larger number of nodes to be compromised. For example, Truebit \cite{truebit} is believed to be the first scalable off-chain computation protocol designed for the Ethereum blockchain. It employs incentive models as well as proofs via off-chain solvers and challengers. If there is a dispute, solvers and challengers employ an off-chain verification that is done through checking the computation steps powered by an on-chain interpreter recursively to reach a point where they disagree with the state change, and the final value is decided on-chain verifying the validity of one of the state changes. Truebit incentivizes challengers via jackpot repository for auditing purposes; outsourcing computation off-chain while maintaining verification on-chain. 

\subsection{Prior Research on Blockchain Oracles}

Although the given classification may provide information about the oracle's role, it does not provide \emph{technical} aspects of blockchain oracles. Xu et al., \cite{orc33,orc15} provide insights about the oracle roles, benefits, and drawbacks from another perspective. They argue that oracles can be implemented as smart contracts in the blockchain network where an external state is periodically injected into the oracle by an off-chain injector. This type of oracle imposes drawbacks on the blockchain as all participants involved in the transaction should trust the oracle, and injected external states cannot be fully verified by other validators, i.e., miners. 

The authors state \emph{reverse oracles} are also in need as sometimes off-chain components may need to have access to data stored on the blockchain (or the smart contracts running on the blockchain) to provide data or checks. In their point of view, one important aspect of the reverse oracle is interactions that should be \emph{non-intrusive}. A non-intrusive interaction is defined as not changing the system core design while it should be through the configuration of smart contracts function or visibility of the transaction on the blockchain. However, adding such a component in a non-intrusive may not be possible, e.g., the Nakamoto consensus algorithm may be inconsistent with normal transaction semantics in enterprise systems. Finally, they explain that a bidirectional binding can exist between off-chain legal contracts and on-chain smart contracts. Digitizing legal contracts and smart contracts could be done on the blockchain where some conditions are implemented by smart contracts. This model comes with drawbacks such as expressiveness since some items of a legal contract cannot be translated into the code. Also, by using a public blockchain, enforceability would be questioned, and different interpretations from the conditions and coding them into the smart contract may exist.

Furthermore, Xu et al., in another study \cite{orc22} consider the blockchain as a software connector, which could be a decentralized solution to centralized shared data storage. Although information transparency and traceability become improved, it increases communication latency due to the mining mechanism resulting in poor user experience. The authors propose that a good practice for public blockchain is to keep the raw data stored off-chain and only meta-data be injected into the blockchain.   

In contrast, Muhlberger et al., \cite{orc21} examine blockchain oracles based on four different scenarios with respect to the data flow direction and the initiator of the data. The inbound oracle data fetches data from the outside world and pushes data onto the blockchain network, and based on the data initiator it can be pull-based or push-based inbound oracle. The former, upon receiving the request, collects the state from off-chain components, and sends the result back to the blockchain (via a transaction). In the latter, the off-chain state is sent to the on-chain component by the off-chain component. The outbound mode is where information from the blockchain is transmitted to the external world. If the outbound becomes pull-based, the off-chain component retrieves on-chain states from an on-chain component. In the push-based, the on-chain component sends the off-chain state to an off-chain component. Muhlberger et al., through quantitative analysis, reveal that the pull-based inbound oracle is the fastest, and the push-based outbound oracle is the slowest. In fact, they explain that while in the inbound pull-based the state is transparent and requests are initiated on-chain, the response time depends on the network speed that causes a bottleneck. In the inbound push-based, data manipulation can happen as the oracle is not deployed or initiated on the blockchain. Contrary, the outbound pull-based oracle may take some time due to the size of the network and requested information, however, in the outbound push-based the network speed or error that occurred in the monitoring process of the blockchain can affect the oracle. 

Beniiche \cite{orc31} reviews the most widely used oracle services such as Provable and ChainLink, and provides the general architectures of the oracles. Beniiche also considers human oracles with an introduction to Augur and Gnosis as the leading prediction markets, and with respect to the discussed architectures, classifies oracles into three design groups; publish-subscribe, immediate read, and request-response. In contrast, Al-Breiki et al., \cite{orc32} study leading blockchain oracles (and services) in terms of trust. The authors review their system architectures along with the advantages and disadvantages. Mammadzada et al., \cite{orc35} present a blockchain oracle framework that assists developers and decision-makers with the design of blockchain-based applications. The framework considers data origins, how data is processed during transactions, validation, and integration to the applications as the fundamental criteria for the framework. 

In comparison to studies \cite{orc21,orc31, orc32, orc33, orc35}, Heiss et al., \cite{orc23} provide a set of key requirements for trustworthy data on-chain, explaining the challenges and the solutions for them. They argue that in addition to safety (avoid triggering blockchain state transition by incorrect data) and liveness (blocking blockchain state transition when data is unavailable) as the characteristics of a distributed systems, \emph{truthfulness} is necessary as it does not allow execution of blockchain state transition by untruthful data provisioning. Based on these properties, there are challenges defined for each of them as; availability, correctness, and incentive compatibility. Incentive compatibility consists of two key characteristics; (1) attributability referred to as mapping data to the source provider, and with respect to the behavior, the data source can be rewarded or penalized, and (2) accountability defined as depositing stake before providing data, and upon the truthful data provisioning, it is paid back. The correctness consists of authenticity and integrity such that the former deals with approving the data source and the latter shows the data should become untampered during the transition, respectively. Finally, the liveness refers to availability and accessibility such that the former implies the availability of the system should be as good as of its least available component, hence, the outage should be kept minimum. The latter means that data must be accessible at any time.

Table \ref{tbl:litrw} provides the summary of existing literature review on the blockchain oracles.

\begin{table}[t!]
\renewcommand\arraystretch{1.35}
\centering
\caption{Summary of existing literature review on blockchain oracle}
\begin{adjustbox}{width=\columnwidth}
\linespread{1}\selectfont
		\begin{tabular}{|p{1.5cm}<{\centering}|p{6.5cm}<{\centering}|p{10cm}}
		\hline \bf{Literature} &
		\bf{Key Research Outcomes}\\
				\hline
		Xu et al., \cite{orc22} & Providing a discussion on validation strategies for oracles which can be internal or external. The former discusses injecting the external state into the blockchain causing latency and trust management issues but the latter's issue is trusted third parties.\\ \hline
		Muhlberger et al., \cite{orc21} & Classifying oracles into four groups based on the information flow direction and data initiator; inbound pull-based, inbound push-based, outbound pull-based, and outbound push-based.   \\ \hline
        Heiss et al., \cite{orc23} & Data on-chain trustworthy requirements and challenges are explained. \\ \hline%data on-chain trustworthy key requirements, and discussing their challenges. \\ \hline
        Beniiche \cite{orc31} & Describing widely used oracles, and human oracles, and with respect to their architecture classifies oracles into three groups; publish-subscribe, immediate read, and request-response. \\ \hline
        Al-Breiki et al., \cite{orc32} & Studying trust in the leading blockchain oracles, and reviewing their system architecture, advantages, and disadvantages. \\ \hline
        Mammadzadea et al., \cite{orc35} &  Describing blockchain oracle framework for assisting developers with the design of blockchain-based applications taking into account data origins, processing data during transactions, validation and integration to the applications.\\ \hline
        Xu et al., \cite{orc33,orc15} & Classifying oracles into three groups; conventional oracles, reverse-oracles, and legal and smart contract pair.\\\hline
% 		\hline
    \end{tabular}
    \end{adjustbox}
    \label{tbl:litrw}
\end{table}

\section{Voting-based Oracles}
\label{sec:voting_oracle}
\begin{figure}[!b]
    \centering
    \includegraphics[keepaspectratio,scale=0.40]{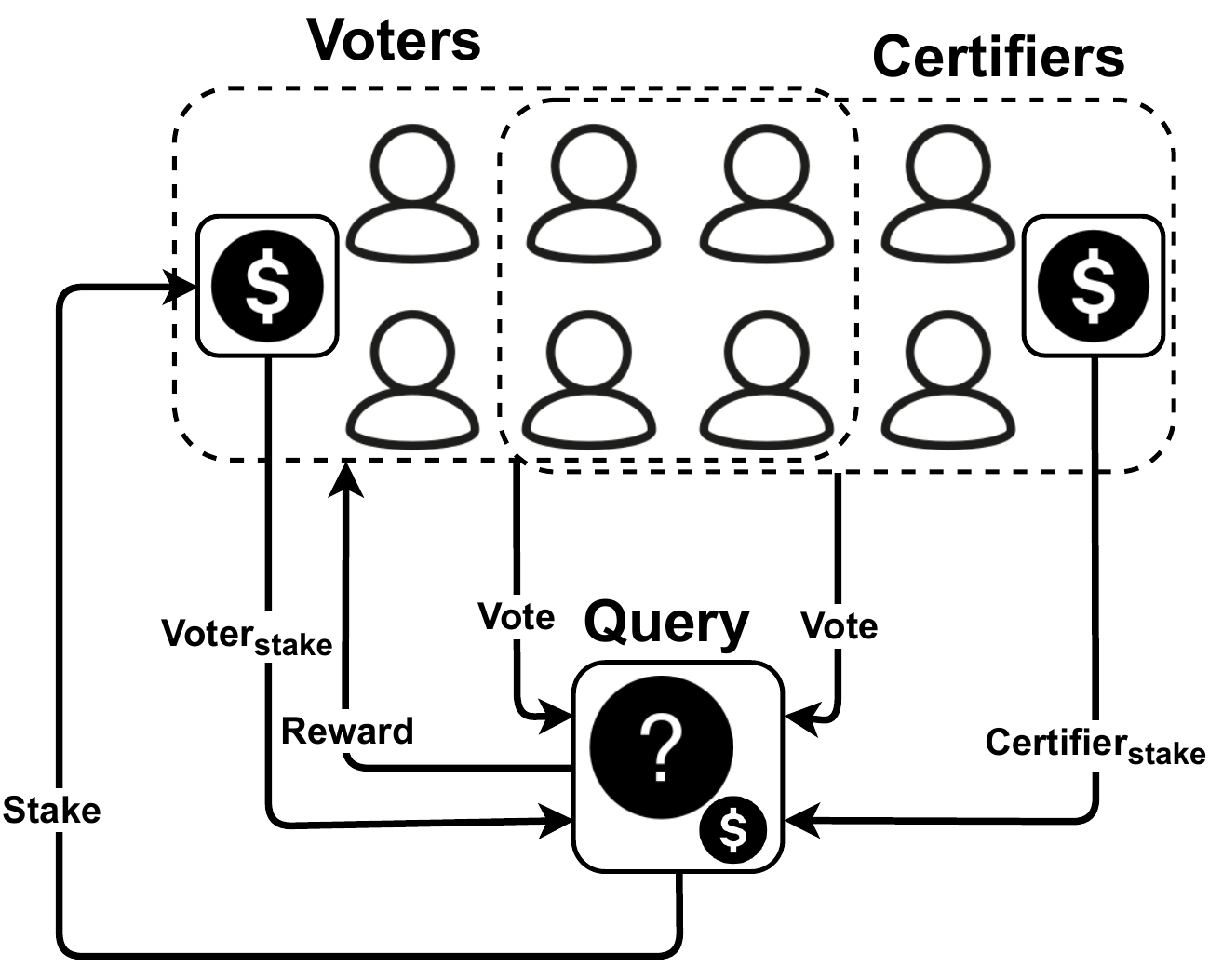}
    \caption{Voting-based oracles overall structure. The reward is only distributed when certifiers and voters' outcomes are matched. Otherwise, certifiers are penalized. Each query comes with a bounty. }
    \label{fig:vote_block}
\end{figure}

Although oracles can provide feedback to the submitted queries, there may be inconsistencies and discrepancies between the received responses from the oracles to the blockchain and smart contracts. To mitigate the issue in terms of the data \emph{correctness}, users can form a set of voters and/or certifiers who are involved in the process of data correctness approval that is shown in Figure \ref{fig:vote_block}. Each voter and certifier put stakes on responses to verify the data. If the outcomes are matched, rewards are distributed between them, otherwise, they are penalized. It may also inherit the game theory concept Nash Equilibrium (e.g., \cite{orc1, orc24,orc14}) defined as the determination of the optimal solution in a non-cooperative game in which each player does not have any incentive to alter the initial strategy. It leads to gaining nothing from changing their initially chosen strategy if other players keep their strategies unchanged. 

Use cases of voting strategies can be seen in prediction market platforms, e.g., Augur, Gnosis, and X Predict Market \cite{ocr3,orc6,xpredict}. Prediction markets are platforms where financial shares in outcomes or facts are created, shared, and exchanged by participants. These platforms enable users to bet on anything, e.g., political forecasting, and receive compensation or become penalized if they are correct or wrong. Prediction markets not only are resistant to manipulation but also are largely scalable, and can help with the aggregation and distribution of unlimited information. Data in the prediction markets depends on the number of participants to take part in because the more participants, the more data and consequently, the more effective the prediction markets are. Prediction markets can be based on distributed oracles, e.g., the Delphi-based prediction market called Omphalos \cite{orc7}, and markets should have a tradable market price at all times known as the market liquidity. Prediction markets can also be multi-dimensional markets in which users not only trade on the state probabilities but also the relationship between dimensions.

The voting-based strategy arises issues in the incentivized platform. There is a term called lazy equilibrium- a form of verifier's dilemma- in which voters always return the same answer to questions to secure profits without performing works for correctness. The other issue is Sybil attacks defined as when attackers out-vote honest nodes on the network by creating multiple fake nodes to take over the network. A Sybil attacker can also employ mirroring that makes oracles to post individual responses based on a single data-source query. Freeloading is another issue and defined as a cheating oracle obtains and copies the response of another oracle without paying per-query fees. This issue threatens the response time of oracles but can be addressed with commit and reveal strategy; sending cryptographic commitments to the responses and in the next round revealing the responses. 

\subsection{Stake-based Systems}
\label{sec:stake_sys}

Astraea \cite{orc1} is a general-purpose decentralized oracle running on the public ledger that relies on a voting-based game strategy. This framework employs monetary and staking fees which assure the system is immune to Sybil attacks. This oracle has entities that may have one or more roles such as submitters, voters, and certifiers who may freely join or exit the system. Boolean propositions are submitted to the system based upon paying a fee by submitters, and voters play a low-risk/low-reward game by placing a small stake of their confidence on the truth of random propositions. In contrast, certifiers play a high-risk/high-reward game by placing a large stake in the outcome of the selected voting and certification process. The outcome of voting and certification is the stake-weighted sum of votes or certifies, respectively, and due to the random nature, it is resistant to manipulation. In weighted votes, the weight (and reward) is affected by the level of the deposit made as the higher the deposit, the heavier the weight, and the higher rewards and penalties. If the outcome of voters and certifiers are matched, they are rewarded, otherwise, the players who take the opposite position are penalized. Hence, this oracle encourages players to place bet/vote on the propositions that have such a high level of confidence. The voting and certification deposit should be small or large relative to the total voting stake on the proposition as the former could not control the outcome and the latter could be penalized and could not tamper with the outcome.

% (when verifier is not willing to perform for verifying correctness of some work)
Kamiya presents Shintaku \cite{orc24} as an end-to-end decentralized oracle that is blockchain-agnostic for deciding on the outcome of binary propositions. It relies on a stake-based voting scheme where voters are rewarded for being honest. The work is an extension to Astraea \cite{orc1}, in which the verifier's dilemma is handled. The issue with the main contribution (Astraea \cite{orc1}) is that the reward pool is non-zero most of the time which is filled in with the penalties obtained from the system. Hence, the system could \emph{lazily} vote and certify toward a single outcome forever. To eliminate this, Kamiya argues that the payout must be zero. In the system, submitters and voters exist, and each voter receives two propositions, and they are only rewarded when their choice for the propositions differs which would lead to returning their bond. To have such a decentralized implementation, the author suggests that voting pools (similar to mining pools but cheaper) can be constructed via an off-chain frontend. It makes on-chain transactions, and voters can freely move between the pools to reduce the risk of centralization. 

Similar to Shintaku \cite{orc24}, Merlini et al., \cite{orc25} also present a paired-question oracle protocol to extract true answers from the public showing a Nash equilibrium of truthful reporting with the advantage of re-balancing. The user submits a pair of antithetic questions with a bond, and the voting users answer them to obtain a reward. The oracle collects votes and checks whether the two questions converged to different answers; if so, the submitter regains their bond, and voters are rewarded (penalized) for agreement (disagreement) with the majority answer. Otherwise, the submitter loses the bond, and voters receive nothing. In comparison to \cite{orc24, orc1}, truthful voters receive larger expected payoffs. 

Cai et al., \cite{orc26} present a peer prediction-based protocol with a non-linear stake scaling for decentralized oracles. In comparison to \cite{orc25}, a light-weight scoring rule controls the rewards for voters, and it considers the behaviors of the other voters with respect to their answers. In addition, the voting weight and award weight with respect to the submitted stake are sub-linear and super-linear scaled, respectively. Questions to voters are assigned by the oracle, and reports in the form of a binary answer including a popularity prediction are collected. The majority of the information determines the oracle answer that is weighted by the associated stakes and adjusted by a sub-linear function. Then, to each report, a score is assigned based on the accuracy and degree of agreement with peers, and only top-scored voters are awarded while the share of award is determined by their stake adjusted by a super-linear function. In comparison to Astraea \cite{orc1}, the system encourages minority voters to vote based on their true opinion to receive an award. The next benefit of the approach is non-linear stake scaling such that an honest voter is incentivized to stake more onto a single report while increases the penalty for a participant to bias the outcome with Sybil attacks. 

Nelaturu et al., \cite{orc14} propose a voting-based game oracle that evaluates the truth or falsity of a query that is similar to Astraea \cite{orc1}. This framework leverages the crowd-sourced voting mechanism that is agnostic to the blockchain consensus protocol and is deployed on the existing platforms such as Bitcoin and Ethereum. There are user roles; submitters who send the queries to the blockchain in conjunction with funding it, randomly selected reporters playing a low-risk/low-reward game, and upon participation stakes must be deposited, and certifiers playing high-risk/high-reward game who have the choice to choose the query they want to put the deposit into. Both voters and certifiers have outcomes defined as a function- the sum of the votes weighted by the deposits. The termination takes place when the query has attached sufficient funds. The author based on the proposed protocol presents a light version of the protocol where only submitters and reporters exist. 

%(which is Turing-complete)
Band protocol \cite{bandchain} is a blockchain-agnostic framework that has its native token for connecting public blockchains to the off-chain information. This framework supports generic data requests and on-chain aggregation with WebAssembly. For retrieving data from Bandchain's oracles, an oracle script is necessary that is defined as an executable program that encodes raw data requests and aggregates final results. Participants in the framework are validators and delegators such that the former is based on a random selection responsible for proposing and committing new blocks to the blockchain. They take part in the consensus protocol by broadcasting votes as well as supporting external data queries. The latter stakes their holding on the network of validators and they can take part in the network governance as their voting power is proportional to the size of the stake they hold. Each oracle script requires outlining the data sources and sending the request to the chain's validators for fetching data. Also, it aggregates the returned results into final results such that the aggregation policy is controlled by the creator. Validators have also voting power in which the tallied voting should be greater than two-third of the system. Upon storing onto the BandChain, an oracle data proof is created.

Razor \cite{orc4} is a decentralized oracle network to offer maximum game-theoretical security without compromising speed. Razor network consists of stakers who are responsible for responding to the queries from a queue, fetching the information from the real world, and are rewarded for reporting honestly. Razor employs proof-of-stake and has its token named Razor used by stakers whose stake amount can influence the network. Razor protocol relies on high economic security, hence, it uses a proof-of-stake chain with Honey Badger BFT as a consensus algorithm network where a large number of individual stakers can participate, and values are reported in consensus with the majority of stakers. Razor uses Median Absolute Deviation (MAD) for measuring the consensus and based on that, votes with absolute division higher than that are penalized. Leveraging the proof-of-stake consensus protocol reduces malicious behaviors for reporting incorrect and inaccurate data points to influence the result. Razor also employs game-theoretical and cryptographic strategies such as a commit-reveal scheme to provide further collusion and censorship resistance. The Razor architecture consists of four parts; oracles composed of stakers for processing queries, job manager that is responsible for accepting and prioritizing queries based on fees, client applications or smart contracts, and users. For each query in addition to fees for using oracles, there is a validity bond incentivizing clients for providing valid and reliable sources and is equal to the maximum potential lost due to the incorrect source. Providing a reliable source by users can be another point of failure and data integrity issue. 

Oraichain \cite{oraichain} is recognized as a data oracle platform that employs artificial intelligence models and uses the ORAI token for payments and governance. The Oraichain core technology is similar to Tellor \cite{tellor} or DIA \cite{dianetwork} but it focuses more on the AI. The issue with the use of AI in smart contracts roots in the languages such as Solidity or Vyper \cite{vyper} which are not suitable for AI model implementation, hence, Oraichain tries to bridge the gap by enabling secure access between smart contracts and AI. Oraichain is a public blockchain that employs a delegated proof of stake (dPoS) consensus protocol providing fast transaction times and completion of data requests quickly. AI models in Oraichain are constantly tested for quality, and per each data request test cases (e.g., face authentication) they come with, and the AI provider must pass these test cases before receiving any payment for sourcing the data request. These test cases are the incentive for providers to keep their AI models more accurate. ORAI token holders by staking their tokens can take part in securing the network and can be rewarded. In fact, Oraichain is a community-driven platform in which ORAI tokens give holders voting power.%, e.g., use cases such as face authentication or detecting fake news.  

\begin{table*}[htbp]
\renewcommand\arraystretch{1.35}
\centering
\caption{The summary of stake-based studies for voting-based oracles}
\small
% \begin{adjustbox}{width=\linewidth}
\linespread{1}\selectfont
		\begin{tabular}{|p{1.5cm}<{\centering}|p{10cm}<{\centering}|p{1.5cm}<{\centering}|p{1.5cm}<{\centering}|p{1.5cm}<{\centering}|}
		\hline \bf{Literature} &
		\bf{Key Research Outcomes} &
		\bf{Query Type} &
		\bf{Sybil Attacks} &
		\bf{Verifier's dilemma}\\
		\hline
		Merlini et al., \cite{orc25} & Presenting a paired-question oracle in which two antithetic questions are submitted. If answers are matched, the  submitter  regains  their  bond  and  voters  are  rewarded(penalized)  for  agreement  (disagreement)  with  the  majority of answers.   & Binary & \checkmark& \checkmark \\\hline
		Adler et al., \cite{orc1} & Presenting a general-purpose decentralized oracle referred to as Astraea. There are different entities such as submitter, voter, and certifier each of which holds stake. Voters play a low-risk/low-reward game while certifiers play the high-risk/high-reward one. If outcomes of voters and certifiers are matched, rewards are distributed. Otherwise, they are penalized. & Binary & \checkmark & Partially \\\hline
		 Kamiya \cite{orc24} & An extension to Astraea, in which two propositions are submitted, and based on different responses to the propositions rewards are distributed. & Binary & \checkmark & \checkmark \\\hline
		 Nelaturu et al., \cite{orc14} & Presenting a framework based on the crowd-sourced voting mechanism employing two strategies for oracles; a version similar to \cite{orc1}, and a light version in which only voters (reporters) exist in the system.  & Binary & \checkmark & \checkmark\\\hline
		 Cai et al., \cite{orc26} & Presenting a peer prediction-based protocol with non-linear stake scaling. It leverages a light-weight scoring rule for controlling rewards for the voters. A score is assigned to each report and based on the accuracy and degree of agreement with peers, the top-scored voters are awarded. & Binary & \checkmark & \checkmark\\\hline
		 Band protocol \cite{bandchain} & A blockchain-agnostic framework with a native token in which validators and delegators are the main participants, the former broadcasts votes and the latter stakes their holdings on the validators governing the network. & (Non)Binary, scalar, categorical&\checkmark&\checkmark \\\hline
		 Razor \cite{orc4} &A decentralized oracle framework that employs a weighted-voting mechanism with respect to the stakers' stake. It uses median absolute deviation for measuring the consensus; votes with absolute deviation higher than the value are penalized.&Categorical or scalar&\checkmark&Partially\\\hline
		 Oraichain \cite{oraichain} &A community-driven platform that provides a connection between the smart contracts and AI models utilizing ORAI token for the platform governance and voting power. & (Non)binary&\checkmark&\checkmark
		\\\hline
		 Synthetic \cite{synthetix} & An Ethereum-based protocol issuing synthetic assets that maintains its token SNX employing decentralized oracles for price discovery. It utilizes Chainlink's independent nodes for price feeds after screening them for security. &Scalar&\checkmark&\checkmark\\		\hline
		Kylin Network \cite{kylinnetwork} & A low-cost cross-chain decentralized oracle based on Polkadot platform that holds the token KYL for operating as an oracle node through staking and to access private data APIs.&Categorical or scalar&\checkmark&\checkmark
		\\		\hline
    \end{tabular}
    % \end{adjustbox}
    \label{tbl:vote_stake}
\end{table*}

Synthetic \cite{synthetix} as an Ethereum-based protocol issues synthetic assets and owns the network token named the Synthetix Network Token (SNX). Synthetic assets are in the form of ERC-20 smart contracts \cite{erc20} known as ``Synth'' that are financial instruments that can be held and tracked without holding them. Decentralized oracles for price discovery of the assets are employed in Synths. SNX holders are encouraged to stake their tokens as they can be paid a pro-rata portion of generated fees by activities on the exchange and contribution to the network. SNX tokens support all Synths, and they are minted when SNX holders stake their SNX as collateral and SNX stakers incur debt. Commercial application program interfaces (APIs) are employed in Synthetix for price feeds of five Synths categories; fiat currencies (e.g., sUSD), commodities (e.g., synthetic gold), cryptocurrencies (e.g., sBTC), inverse cryptocurrencies, and cryptocurrency indexes \cite{inversesynth}. Prices from decentralized sources are averaged to aggregate the final value for each asset, and these price feeds are supported by both Chainlink’s independent node operators and Synthetix. These independent Chainlink nodes are employed when they are reviewed in terms of security and having a proven track of successful record of providing data. The price that exists on-chain is updated with respect to a price deviation model and a minimum time-based update. The on-chain prices are also updated every 1\% change from the previous price, or a minimum of once every hour even if volatility is low. 

Kylin Network \cite{kylinnetwork} employs the Polkadot platform \cite{polkadot} to create a cross-chain decentralized oracle network at a low cost. It holds the native token KYL, and provides the application, blockchain, or parachain of any form access to the external data, and provides a wide variety of data feeds, e.g., weather or stock market through connecting to APIs. This token assists the on-chain governance and keeps the platform decentralized as it develops. KYL token is also a requirement (through staking) for operating as an oracle node or for opening a dispute. Additionally, KYL is used for payment to access private data APIs. In the Kylin network there are four major components; (1) analytics for improving the efficiency of applications, (2) a query engine for the public and API access, (3) data oracle as a decentralized data feeding protocol powered by Polkadot, and (4) marketplace as an open platform for pricing and trading data. Kylin network employs a network of data providers, oracle nodes, and arbitration nodes to keep the data sourcing decentralized. 

Table \ref{tbl:vote_stake} provides a summary of studies that employ stake-based strategy for finalizing the oracle outcome. The use of stakes on the outcomes can mitigate the Sybil attack in the presence or absence of a general consensus algorithm, however, it may be prone to the verifier's dilemma issue.  

\subsection{Multi-signature based Systems}
\label{sec:ms_sys}
Orisi \cite{orc12} is a Bitcoin-based distributed system for the creation of oracle sets run by independent and trustworthy parties. The Orisi aims to mitigate flaws and issues arising from a single (server) oracle, i.e., the point of failure. In this framework, a majority of oracles need to agree on the outcome for a transaction to be finalized as it would be very expensive and hard to bribe more than half of the oracles. For this purpose, Orisi leverages multi-signature addresses (oracles and sender/receiver) such that the money from senders and receivers are placed into the addresses (i.e., safe address). 
A multi-signature address is defined as an address on the blockchain associated with more than one private key, and a multi-signature transaction needs to have more than one private key for transaction authorization. They are considered as m-out-of-n addresses requiring m keys out of a total of n keys to sign a transaction for adding into the blockchain. To increase the security, Orisi uses Bitmessage that is a trustless decentralized peer-to-peer protocol for sending and receiving messages securely \cite{bitmessage}. The Bitmessage protocol employs a hash of the public key and has a message transfer mechanism similar to Bitcoin transactions such that each message requires proof-of-work. Messages are broadcast and each recipient should apply its private key to decode them. Hence, employing Bitmessage protects IP addresses for communication with oracles, and senders use Bitmessage to broadcast the transaction on the network. Oracles check the validity of transaction and rules, and then the sender and receiver, upon realizing oracles' acknowledgments on the transaction validity, send the fund to the ``safe''. Once oracles notice the condition, they add their signature to the transactions that are broadcast to the network. Also, Orisi uses a timelock verdict when the source becomes unavailable, and it leverages dedicated oracle data feeds or mediation protocol for mitigating hacks. Mediation protocol is defined within a time frame, the receiver can challenge the verdict and a human operator delivers arbitration.

Gnosis \cite{orc6} is an open-source infrastructure for building prediction markets on the Ethereum platform aggregating relevant information from human and artificial intelligence agents. The outcome of events is exchanged in the prediction markets, and Gnosis provides the ability to trade cryptocurrencies represented as the outcome of events on the platform which can be categorical or scalar. Gnosis consists of three primary layers; (1) Core Layer which interacts with Ethereum blockchain providing the base functionalities for event contracts that monitor and set the outcome token creation and settlement, and a market mechanism, (2) Service Layer that offers optimization tools such as chatbots and stablecoins, and (3) Application Layer that is the Gnosis frontend and targets a particular prediction market or customer segment. Third-party applications in the layer may charge additional fees or use alternative business models, e.g., market making, information selling, or advertising. Consensus can be done via voting, i.e., it requires multiple signatures for approval. Oracles can be on-chain, centralized, and the ultimate oracle is triggered by staking 100 Ethereum if users disagree with the reported value.

\begin{table*}[htbp]
\renewcommand\arraystretch{1.35}
\centering
\caption{The summary of Multi-signature based studies for reputation-based oracles}
\small
% \begin{adjustbox}{width=\linewidth}
\linespread{1}\selectfont
		\begin{tabular}{|p{1.5cm}<{\centering}|p{10cm}<{\centering}|p{1.5cm}<{\centering}|p{1.5cm}<{\centering}|p{1.5cm}<{\centering}|}
		\hline \bf{Literature} &
		\bf{Key Research Outcomes} &
		\bf{Query Type} &
		\bf{Sybil Attacks} &
		\bf{Verifier's dilemma}\\
		\hline
		Orisi \cite{orc12} & A bitcoin-based distributed system for creating a set of oracles run by independent and truth-worthy parties leveraging multi-signature addresses, and employs Bitmessage protocol for sending and receiving messages securely.&Non-binary& \checkmark& None\\\hline
		 Gnosis \cite{orc6}&An open-source infrastructure to build prediction markets on the Ethererum platform. Event outcomes are traded in the prediction markets, and Gnosis enables trading cryptocurrencies represented as the event outcomes on the platform. & Categorical or scalar&Partially&Partially\\\hline
		
		Delphi \cite{orc7} &Offering a light-weight strategy that  employs a weighted signature framework  called  Pythia and a compound token. Distributed  oracles rely on multi-signature contracts to generate the oracle.  &Non-binary& \checkmark& \checkmark\\ \hline
		
		Moudoud et al., \cite{iotbc} & Employing an oracle network for data veracity where follows m-out-of-n multi-signature transaction should be reached for the consensus.  & Non-binary & Partially & Partially\\\hline
		
		DOS Network \cite{dosnetwork} & A layer-2 protocol provides off-chain computation in a decentralized way, and has two partitions on-chain and off-chain as a client software for implementation of the core protocol employing m-out-of-n multi-signature transaction for the consensus.  & Non-binary & \checkmark & Partially\\		\hline
    \end{tabular}
    % \end{adjustbox}
    \label{tbl:vote_multisignature}
\end{table*}

Delphi \cite{orc7} offers a light-client strategy in which event filters and social application functions are bundled in helping users to build and deploy distributed oracles. Delphi employs a weighted signature framework called Pythia that leads to faster input arbitration, understandable oracle interfaces for developers, and providing flexibility and extensibility. These distributed oracles can be resistant to Sybil attacks, and rely on multi-signature contracts highlighting authorization from more than one entity. It is necessary to generate the oracle output with respect to the weights and threshold which makes the consensus trivial. This platform also allows to re-weight signatures, e.g., to vote out misbehaving oracles, and a decaying weight strategy may be applied to the involved oracles such that the weight replenishment requires being honest for providing the truth. The platform leverages a compound token; (1) a minimal token that is atomic, lightweight, easy to use and understandable, and compatible with existing token solutions. (2) The signal component gives the rich functionality for market signaling, seamless and permissionless feature upgrades, and tracking values or rankings over time, and making it suitable for voting leading to providing a Sybil attack resistant mechanism based on the coin. Finally, (3) the trustless peg component unites two tokens into a single token architecture that gives users toggling balance freely.

Moudoud et al., \cite{iotbc} present a permissioned and light-weight blockchain architecture for supply chain use case consisting of distributed internet of things (IoT) entities. It has two blockchains; private for storing private information and the public for tracking produce and providing general information to the public. It is a peer-to-peer overlay network involving supply-chain members identified by a public key and any new member is added when the minimal number of members reached an agreement. Since data is collected from different locations, oracles are employed to check the correctness of data, hence, the proposed blockchain uses multiple oracles- the oracle network- for data veracity approval to be divided and approved by multiple parties. Due to the limited block size, data is stored off-chain, and the metadata is kept on-chain. The consensus used for the oracle follows m-out-of-n multi-signature transactions that should be reached among oracle parties.

DOS Network \cite{dosnetwork} as a layer-2 protocol provides off-chain computation in a decentralized way to feed the results to the blockchain. A layer-2 protocol is defined as a secondary framework or protocol being built on top of an existing blockchain. Node operators are incentivized by DOS token for providing honest services to receive rewards in addition to providing unlimited decentralized verifiable computation oracle to mainstream blockchains. DOS network is resistant to Sybil attacks and is chain-agnostic (i.e., it can deal with any smart contract platform) and also is horizontally scalable offering more capability and computation when more nodes run the DOS client software. DOS network consists of two partitions; (1) on-chain for providing a variety of functionalities, and (2) off-chain as client software for implementation of the core protocol. The latter is used by third-party users to obtain economic rewards and constitute a distributed network. The consensus among off-chain clients in the DOS network is achieved through employing techniques such as unbiased verifiable randomness generation and non-interactive and deterministic threshold signatures. Computation oracles are equipped with zkSNARK \cite{zksnarks} enabling decentralized computation marketplace for commercial computation applications monopolized by tech giants (e.g., like video/audio transcoding or machine learning model training). Upon availability of a query, randomly selected nodes reach consensus by the t-out-of-n threshold signature algorithm (in addition to verifiable random function (VRF)). The agreed result is reported back to the DOS on-chain system, as long as more than $t$ members are honest where nodes' identity and quality of service (QoS) (e.g., responsiveness/correctness) performance are recorded on-chain for monitoring and data analysis purposes. Since the honest nodes earn an even split of the payout, the DOS network is also protected against freeloading. Each node to join the network needs to deposit DOS token mitigating Sybil attacks and enhancing security. The DOS token is natively supported for payment, and an extra payment for stablecoins (e.g., USD coin) exists.

Table \ref{tbl:vote_multisignature} shows a summary of the studies based on the multi-signature strategy.

\subsection{Schelling point Systems}
\label{sec:sch_sys}
Buterin \cite{schellingcoin} presents a mechanism that relies on the concept Schellingcoin for the creation of a decentralized data feed. The mechanism works as follow; users submit a hash of data (e.g., price feeds) along with their Ethereum address, and in the block, after users provide the value (plus assigning a deposit to it), the submitted values are sorted and each user whose submitted value is correct and is between the 25th and 75th percentile is rewarded. In other words, deposits are reassigned in a way that reported values that are far from the median are penalized while values that are closer to the median are rewarded. The mechanism is not immune to Sybil attacks but proof-of-work or proof-of-stake mechanisms can be used for this purpose. There is also a limitation to this approach because if an entity controls more than 50\% of the votes, the median can be set to any wanted value. The other issue is micro-cheating that is defined when slight changes are applied to the value frequently, and participants can slightly tweak their answer toward one direction and thereby pushing the median toward their desired point. This can be addressed in a centralized way, e.g., defining a value unambiguously, or a coarse-grained approach for the value to mitigate the slight changes. 

Usage of the median point can be seen in \cite{mdao} that is known as Maker Protocol or the Multi-Collateral Dai (MCD) system built on the Ethereum blockchain for the creation of currency. One element of the system is oracles (assumed trusted and approved) being responsible for a real-time market price of the collateral assets. These oracles are decentralized and have independent individual nodes called Oracle Feeds. Oracles have a security module and medianizer (a smart-contract for collecting price data from Feeds and providing a reference price by a median), and each oracle feed has a tool called Setzer for pulling median exchange prices and pushing them to a secure network (i.e., database protocol) where relayers aggregate the price data. Medianizer receives a transaction from the relayers and determines the median of the reported values and publishes it as a queued reference price which is delayed by the Oracle Security Module.

The Oracul system \cite{oracul} is also based on the SchellingCoin concept that has been introduced by Vitalik. In this system, a $\delta$ is considered that represents a spread tolerance range for the reported value. If the $\delta$ is zero, every reported value except the median one is penalized with respect to its distance from the median. When $\delta$ is bigger than zero, all the reported values are valid and will receive a share of penalties produced based on the reported values outside of the range. 

Table \ref{tbl:vote_schelling} presents a summary of the studies based on the Schelling point concept, and it is understood they are not completely resilient to Sybil attacks.

\begin{table*}
\renewcommand\arraystretch{1.35}
\centering
\caption{The summary of Schelling point based studies for voting-based oracles}
\small
% \begin{adjustbox}{width=\linewidth}
\linespread{1}\selectfont
		\begin{tabular}{|p{1.5cm}<{\centering}|p{10cm}<{\centering}|p{1.5cm}<{\centering}|p{1.5cm}<{\centering}|p{1.5cm}<{\centering}|}
		\hline \bf{Literature} &
		\bf{Key Research Outcomes} &
		\bf{Query Type} &
		\bf{Sybil Attacks} &
		\bf{Verifier's dilemma}\\
		\hline
		Buterin \cite{schellingcoin}&Presenting an Ethereum-based blockchain for the currency creation. It relies on trusted oracles for fetching data feeds relying on the median value for providing a reference point.  &Scalar&\xmark&\xmark\\\hline
        MarkerDAO \cite{mdao}&Presenting a mechanism that relies on the concept of Schellingcoin for the  creation of a decentralized data feed.  &Scalar&\xmark&\xmark\\\hline
        Oracul \cite{oracul}& Utilizing a spread tolerance range for rewarding or penalizing the price reporters.  &Scalar&Partially&\xmark\\\hline
    \end{tabular}
    % \end{adjustbox}
    \label{tbl:vote_schelling}
\end{table*}

\subsection{Token-based Systems}
\label{sec:token_sys}
% \subsubsection{Prediction Markets}
% \label{sec:prdmrkt}

Augur \cite{ocr3} is a decentralized oracle and platform for prediction markets and is believed to be an early prediction market implementation. It was originally designed as an extension to the Bitcoin Core source code employing Bitcoin Script-based logic but later on it switched to the Ethereum smart contract architecture. Users in Augur select the outcomes of events and they hold reputation tokens. Progressively-larger reputation bonds which later on are divided into multiple versions are posted by token holders for disputing the proposed market outcome. The token (REP) is required for the market creators and reporters who stake their REP on a market's outcome which is similar to Truthcoin. If a reporter's outcome does not match with the other reporters', Augur re-distributes its stake on the outcome to other reporters whose outcomes are matched. In the Augur system, the creator of a market posts two bonds; the validity for incentivizing creators for creating well-defined events, and the creator (paid in REP) for choosing a reliable reporter. There is a period for the designated reporter to report the outcome, and if it fails, the bond will be distributed to the first reporter. Upon receiving the tentative report, there is a dispute window such that REP holders may participate in creating a dispute that consists of staking REP on an outcome other than the tentative one. The dispute is resolved successful when the dispute stake on some outcomes meets the dispute bond size for a round. The Augur system functions as a single oracle that leverages an iterative commit-and-reveal process where token holders are free to participate. The collected platform fees are shared among all the voters requiring relatively active participation (e.g., voting and appealing). The system may let voters be settled as long as their ability to collude is minimized. Also, Augur leverages a validation-dispute protocol in which token holders report or challenge the outcome. 

Tellor \cite{tellor} is an Ethereum-based decentralized oracle that employs proof-of-work (resistant to Sybil attack) and fetches any data requests in the Tellor smart contract. In addition to enabling developers to query Tellor's on-chain database for data, Tellor holds a token named the Tributes (TRB) for incentivizing miners not only to provide data legitimately, but also to vote for validation of data in a dispute. The token is used by users for requesting data and to reward miners. Participant miners deposit the token in the Tellor's smart contract and are rewarded or penalized in case of providing correct or incorrect data, respectively. Tellor chooses the first five miners to provide the proof-of-work solution and the five off-chain data points to be rewarded with newly minted tokens and the accumulated tips for the specific data requests. When the same data is requested by other users, they need to pay a ``tip'' to incentivize the miners more, and in an interval-based manner, the Tellor's smart contract picks the most funded query. When values become available, they are sorted and the first five values are selected, of which the median value is saved onto the chain, and miners are rewarded. 

Decentralized Information Asset (DIA) \cite{dianetwork} as an Ethereum-based ecosystem for an open financial ecosystem acts as a bridge between smart contracts on-chain and off-chain data sources in a verifiable and reliable way. DIA employs crypto-economics to incentivize and validate data coming from data providers, in addition to using the community wisdom for validation and data sourcing. DIA has three main building blocks as (1) data collection mechanisms known as scrapers supported by a centralized backend, (2) a flexible database layer for handling all different kinds of data streams, and (3) distribution through REST API and oracles operating on multiple blockchains. Stakeholders for the information unavailable on the DIA blockchain, submit a funded request that becomes public, and the requester pays the bounty in DIA token for the data provision upon validating the information. Scrapers are created by data providers that may be connected to the on-chain smart contracts or APIs for retrieving the requested data, and analysts exist who are responsible for verifying the submitted code by staking mechanisms. In case of incorrect submitted data, the code is challenged via staking DIA tokens, and based on a voting strategy, the DIA community evaluates the right solution and who should be rewarded. The outcome is kept in a database which is an immutable and open-source database and is also published on the DIA platform. In this platform, historical data can be accessed free of charge while specific APIs and live prices are paid by DIA tokens.

Sztorc presents Truthcoin \cite{orc2}, renamed to Hivemind, a blockchain-based platform for prediction markets. Hivemind is a peer-to-peer decentralized oracle protocol that acts as a side chain to Bitcoin and inherits all the Bitcoin assumptions. It provides the ability of multi-factor decision governance in the prediction market and aims at the information aggregation problem with the help of monetary aspect, transparency, and censorship resistance of the blockchain. The platform uses dual tokens in which the Bitcoin (i.e., CashCoins) serves as the interface for the users, and the VoteCoins as the reputation layer indicating a user reputation on the platform. Hivemind can host many oracles named branches (with respect to a topic), each of which holds a set of VoteCoins; the higher VoteCoins percentage in the branch, the higher degree of voting influence. Weighted votes assisted by Votecoins provide the outcomes, and the malicious behavior is controlled by collapsing the coin market value, miner vetoes, and overrides. Owners' VoteCoins are prone to be lost due to refusing to participate in voting or voting differently from the majority. Decisions are resolved by the voters which can be Boolean or scalar leveraging the VoteCoin for a decision on the outcome, similar to the process used in Augur by the use of Reputation tokens (REP). Hivemind applies singular value decomposition (SVD) for outcomes' calculation, and market decisions are divided into branches having their parameters and VoteCoins, and a decision is made for the branch. 

\begin{table*}[htbp]
\renewcommand\arraystretch{1.35}
\centering
\caption{The summary of token-based studies for voting-based oracles}
\small
% \begin{adjustbox}{width=\linewidth}
\linespread{1}\selectfont
		\begin{tabular}{|p{1.5cm}<{\centering}|p{10cm}<{\centering}|p{1.5cm}<{\centering}|p{1.5cm}<{\centering}|p{1.5cm}<{\centering}|}
		\hline \bf{Literature} &
		\bf{Key Research Outcomes} &
		\bf{Query Type} &
		\bf{Sybil Attacks} &
		\bf{Verifier's dilemma}\\
		\hline
		Augur \cite{ocr3} &A decentralized oracle and platform for prediction markets that uses token (REP) for market creators and reporters who stake their REP on a market's outcome.&Non-binary or scalar&\checkmark&Partially\\\hline
		Tellor \cite{tellor}& A decentralized oracle that employs proof-of-work and its native token to return the outcome data based on the median value of first five data providers.& Aany type of data&\checkmark&\checkmark\\\hline
		Decentralised Information Asset (DIA) \cite{dianetwork}& A decentralized oracle that employs crypto-economics to incentivize and validate data, and has three main building blocks such as (1) data collection mechanisms, (2) a flexible database layer, and (3) distribution. Staking and voting mechanisms are used to resolve disputes. &Any type of data&\checkmark&\checkmark\\\hline
		Sztorc \cite{orc2}& A peer-to-peer decentralized oracle protocol acting as a side chain to Bitcoin, employing dual tokens CashCoins to serve as the interface for the users, and the VoteCoins as the reputation layer indicating a user reputation on the platform.&Binary or scalar& \checkmark& Partially\\\hline
		Polkadot \cite{polkadot}& Providing the platform for creation of user-defined blockchains that maintains two blockchains; relay chain and parachain that employs a form of consensus as proof of stake as well as the DOT token for voting. &Non-binary&\checkmark&\checkmark\\\hline
		Mobius \cite{mobius}& A  Stellar  blockchain-based  network assisting developers  with  creation of their  decentralized applications and oracle  systems via series of APIs. It holds MOBI token and employs proof of stake for incentivizing/penalizing users. &Any type of data&\checkmark&\checkmark\\	\hline
		
		Zap Protocol \cite{zapprotocol}& A decentralized oracle and a permissionless protocol, which permits oracles to be built on the protocol as a form of investment via ZAP/DOT tokens meaning the oracle model and the underlying data can be monetized based on supply curve. & (Non)Binary&\checkmark&Partially\\	\hline
    \end{tabular}
    % \end{adjustbox}
    \label{tbl:vote_token}
\end{table*}

Polkadot \cite{polkadot} is software that incentivizes a global network of computers to operate a blockchain on top of user-defined blockchains. Polkadot maintains two types of them; (1) the main network that is called a relay chain on which transactions are permanent, and (2) user-created networks which are referred to as parachains employing a variation of proof-of-stake consensus known as nominated proof of stake (NPoS). One advantage of parachains is their customization for any number of use cases as well as feeding data into the main blockchain, providing the parachain transactions with benefits in the same level of security as the main chain, and keeping the transactions secure and accurate. This only leverages computing resources that are necessary for running the main chain. In addition to the main chain and parachains, the bridge blockchain exists that assists Polkadot network with interacting with other blockchains. There are different roles for those who stake DOT (i.e., the native token) as they can be validators for voting and validation of data. Also, they can be nominators for the selection of trustworthy validators, and collectors that are responsible for storing the history of each parachain and aggregating parachain transaction data into blocks, and finally fishermen for monitoring the network to report to validators. DOT token holders in the network can use their coins to prove/reject changes proposed by others to the network.

Mobius \cite{mobius} is a Stellar blockchain-based network \cite{stellar} enabling developers to create their decentralized applications and oracle systems through series of APIs connecting applications to the blockchain. Mobius has a token called MOBI to facilitate transactions and to supplement Mobius protocol (i.e., cross-blockchain standards) to assist payments, logins, and the oracle management through simple APIs and developer frameworks. Mobius employs the proof of stake model requiring participants to stake a certain amount of tokens to be granted the privilege for contribution to the maintenance and growth of the network. There is a Universal Proof of Stake Oracle Protocol for connecting the real-world data to the blockchain, and by staking Mobius tokens, high authenticity data transmissions to the blockchain and high-throughput data transfer to secure smart contracts are achieved. This proof of stake protocol incentivizes/penalizes oracles for providing correct/incorrect data while the Mobius Universal Proof of Stake oversees vesting and staking MOBI token. In addition, a quality threshold for the selection of oracles is used in such a way that separate markets are created, and quality scores are aggregated based on the past performance, staking MOBI, and Proof of Verification for auditing oracles.

Zap Protocol \cite{zapprotocol} is a decentralized oracle and a permissionless protocol that is based on Ethereum with a focus on three main components as data, tokens, and bonding curves that are mechanisms powered by on-chain smart contracts for controlling the direction of a decentralized autonomous organization. By this protocol, in addition to buying/selling data, liquid tokens and pricing curves can be defined providing the ability to (un-)bond money to the curves. This protocol, unlike \cite{tellor}, permits oracles to be built on Zap as a form of investment which means the oracle model and the underlying data can be monetized, and the better the oracles, the more money bonded with them. Thus, users can detect which oracles are more reliable and profitable. Although the Zap protocol employs tokens (i.e., ERC20 \cite{erc20}) on its network, they are defined differently as they can be liquid and the supply can dynamically be adjusted to demand, and these tokens can be traded on decentralized exchanges. Bonding ZAP to the oracle by subscribers (i.e., smart contracts that require data) results in receiving DOT tokens for querying the oracle, and the ZapMarket smart contract helps the exchange of InterPlanetary File System (IPFS) \cite{ipfs} public keys for creation of a private IPFS publish-subscribe channel for pushing data to the subscriber by the oracle. A data provider can be an oracle by being registered with the ZapMarket smart contract that needs to define the DOT/ZAP supply curve determining the distribution of DOT per ZAP.

Table \ref{tbl:vote_token} provides a summary of the studies for token-based systems, and it illustrates that employing a token for the network/protocol governance is of great importance.

\subsection{Conventional Systems}
\label{sec:con_sys}

While stake-based systems are beneficial to be used for oracles, there are studies whose platforms rely on either a single data source, e.g., \cite{velocity} or multiple data sources, e.g., \cite{orc34}, without any mechanism to verify the data integrity and correctness, e.g., \cite{aternity}. Also, data can be directly submitted to the blockchain without a need for third-party data providers or obtaining processed data from a distributed ledger \cite{iota}. One can assume that they are trusted entities whose \emph{vote} for data verification is reliable, however, this reliability comes with flaws such as a single point of failure or tampering. 

Eskandari et al., \cite{velocity} present Velocity as an Ethereum-based decentralized market to trade a custom type of derivative option. It employs a tool called PriceGeth to fetch the price information in real-time. A derivative is defined as a contract between two or more parties, and its values are determined based on the agreed underlying financial assets. The price feed consists of PriceFetcher saving exchange prices into a database at specific intervals, BlockListener for monitoring Ethereum blockchain for new blocks, and a PriceGeth server for sending data to the PriceGeth smart contract for updating the latest price. The PriceGeth interacts with the PriceFetcher module for exchange prices. While Velocity employed Provable (in addition to TLSNotary for authenticity proof- see Section \ref{sec:reputaion}), issues such as delay and the insufficient amount of gas led to presenting PriceGeth for nearly real-time price acquisition. However, the structure of PriceGeth by employing the PriceFetcher module is a point of failure as there is no mechanism to avoid data manipulation while PriceFetcher can be assumed a trusted entity.

Zhang et al., \cite{orc34} extend the industrial IoT (IIoT) framework for providing a solution for trustless data sharing by employing an encrypted ledger for reducing the risk of data tampering. The framework consists of a blockchain controlled by the consensus rule of byzantine fault tolerance and different layers such as IoT, fog, micro-service, and decentralized applications layers. The cloud-based micro-service layer (i.e., APIs) provides data feed to smart contracts in decentralized applications. Micro-service layer acts as Platform-as-a-Service (PaaS) layer in IIoT to provide computational power and hold APIs for smart contracts. The framework uses a fog layer to alleviate the slow response due to the massive data-producing rate at the IoT layer. The framework, upon receiving data requests and approval by the customer, allows IoT data to be acquired by net gates (i.e., in the form of hardware oracles) which is encrypted and anonymized by the user’s private key.

Arts et al., \cite{aternity} present Aternity as an open-source blockchain-based framework that uses proof of work cuckoo cycles (a graph-theoretic and memory-intensive problem for finding cycles in the graph), and leverages state channels as (on)off-chain encrypted peer-to-peer communication for a smart contract execution which may not be recorded on-chain. This allows the framework to use an off-chain contract with an on-chain oracle for providing on-chain data. The framework has Sophia- the smart contract language- and contracts are compiled into bytecode executed on highly efficient virtual machines FATE. The framework has a native token that can be used for coordination between participants as their amount of tokens represent their influence of vote on the system, and it is required for any operations on its blockchain. There are generalized accounts providing flexibility for transaction authentication managed by a smart contract. Upon transaction execution, the authentication function in the smart contract and the account evaluate the authentication data, and if it fails, the transaction is discarded, otherwise, it incurs charges. Moreover, there are register oracle transactions for the announcement of oracles to the chain specifying queries and response format associated with a fee. Publicly available oracles monitor the blockchain for queries, and since the response on the chain is public, it causes privacy issues. 

\begin{table*}
\renewcommand\arraystretch{1.35}
\centering
\caption{The summary of conventional studies for voting-based oracles}
\small
% \begin{adjustbox}{width=\linewidth}
\linespread{1}\selectfont
		\begin{tabular}{|p{1.5cm}<{\centering}|p{10cm}<{\centering}|p{1.5cm}<{\centering}|p{1.5cm}<{\centering}|p{1,5cm}<{\centering}|}
		\hline \bf{Literature} &
		\bf{Key Research Outcomes} &
		\bf{Query Type} &
		\bf{Sybil Attacks} &
		\bf{Verifier's dilemma}\\
		\hline
		IOTA oracle \cite{iota} & Presenting First Party Oracle where data is sent to the IOTA Tangle directly without a need for third-party data providers or a distributed ledger. &Scalar&\xmark&\xmark\\\hline
		Eskandari et al., \cite{velocity} & Presenting Velocity, an Ethereum-based decentralized market for trading a custom type of derivative option (e.g., price feeds). PriceGeth is employed as a tool for fetching the information in a real-time fashion (single data source). &Scalar&\xmark&\xmark\\\hline
		
         Zhang et al., \cite{orc34} & Extending the industrial IoT framework to provide trustless data sharing through an encrypted ledger for reducing the risk of data tampering. The framework consists of multiple layers where the micro-service layer provides data feed to the smart contracts (multiple data sources). &Non-binary &\xmark&\xmark\\\hline
         
         Arts et al., \cite{aternity}& An open-source blockchain-based framework that employs proof of work cuckoo cycles, and uses state channels as (on)off-chain encrypted peer-to-peer communication.&Non-binary&\checkmark&\checkmark\\\hline
         
        Hyperledger Fabric \cite{orc13,hyperledger}& An open-source blockchain but permissioned framework for the enterprise context in which participants' identities are authenticated. The framework has a configurable, modular architecture, and does not need to have a native cryptocurrency. It allows smart contracts (i.e., chaincodes) to run within a container by general-purpose programming languages. &(Non)Binary, scalar, or categorical&\checkmark&\checkmark \\	\hline
        
        Compound protocol \cite{compoundprotocol}& An Ethereum-based system that is recognized as interest markets. It employs cToken as the native token and Open Price Feed as a decentralized price oracle that has entities such as posters, reporters, and view. & Scalar&\checkmark&\checkmark\\\hline
    \end{tabular}
    % \end{adjustbox}
    \label{tbl:vote_conventional}
\end{table*}

Hyperledger Fabric is an open-source blockchain but permissioned framework hosted by the Linux Foundation \cite{orc13,hyperledger} for the enterprise context in which participants' identities are authenticated. The framework has a configurable and modular architecture and does not need to have a native cryptocurrency allowing smart contracts (i.e., chaincodes) to run within a container and to be coded with general-purpose programming languages. It supports customized consensus protocols and benefits industries, e.g., track-and-trace of supply chains, healthcare, banking, or insurance, where data cannot be exposed to unidentified entities, hence, increases privacy. The framework employs execute-order-validate architecture for transactions in which ``execute'' checks the correctness after execution, the ``order'' applies a customized consensus protocol, and ``validate'' determines transactions against an application-specific endorsement policy. In fact, it reveals how many peer nodes and which peer nodes are required to confirm the correctness of smart contract execution. In Hyperledger Fabric, there are channels as a means of the communication channel between members, and these channels add an extra layer of access control improving confidentiality. Participants in these channels establish a sub-network where each member can have access to a particular set of transactions. Chaincodes on the Hyperledger Fabric run on the peers and create transactions. Invocation of chaincodes can lead to update or query the ledger, and based on the proper permission another chaincode can be invoked to access the state in the same or different communication channel. Chaincodes have endorsement policies for the selection of peers to execute the chaincode through checking whether enough endorsements are present, versioning checks are done, and are derived from suitable entities, and then verify the result for making sure that the transaction is valid. Hence, they can be considered as platform-supported (distributed) oracles for the Hyperledger Fabric.

Compound protocol \cite{compoundprotocol} as an Ethereum-based system is recognized to be interest markets allowing borrowers to obtain loans when lenders put their crypto assets into the protocol earning variable interest rates. The protocol has its native token called cToken which is a form of ERC20 token and requires approval to be minted initially. The protocol employs the Open Price Feed as a decentralized price oracle that is built on the Ethereum blockchain. This price feed consists of entities such as Reporters, Posters, and a View that is a set of Reporters used to obtain the finalized prices. Posting/storing price data can be done by any user who has access to a reliable source that is signed by a private key to be available to the public. Posters post the data on the chain, and this responsibility is shared among many Posters. The Compound protocol leverages a View contract employing a single Reporter and verifies all the reported prices within an acceptable time, and in the presence of enough reporters that are approved through governance, a median price can be leveraged. These price feeds via the View contract can be configured for developers while the Reporters in the Compound View contract may require approval by the Compound governance. The Open Price Feed allows price data reported by reporters to be signed via a known public key, and posters that can be any Ethereum address can put the value on the chain. The interest rates paid and received by borrowers and lenders are determined by the supply and demand of each crypto asset.

Table \ref{tbl:vote_conventional} provides a summary of the studies in this category for voting-based strategies.

\section{Reputation-based Oracles}
\label{sec:reputaion}

Different data sources/providers may be employed by oracles, hence, this necessitates to employ evaluation mechanisms for selection, monitoring their truthfulness for the provided data, and the retrieved information is intact. Information retrieval by oracles may necessitate mechanisms to firstly ensure the received data is untampered, and secondly \emph{identify} which oracles have more potential to be trusted for the outcome. The former can be achieved by \emph{authenticity proof} mechanisms attached to the data, while the latter can be managed by a reputation component which is responsible for oracles' evaluation. The reputation-based oracles may be assisted by authenticity proof mechanisms that facilitate verification of the data retrieved from external resources and mainly ensures that the data is untampered and genuine. For this purpose, non-repudiation types exist; non-repudiation of origin, non-repudiation of receipt, and non-repudiation of conversation, each of which provides proofs \cite{tlsn}. Non-repudiation of origin supplies proof that a message comes from a specified originator blocking attempts to deny having sent the message. In contrast, non-repudiation of receipt proves that a message is received by a specified recipient falsifying the recipient's false claim. Finally, non-repudiation of conversation generates proof for the occurrence of a conversation between parties. 

Oracles for retrieving data from sources may employ a secure HTTP connection (i.e., HTTPS) powered by the TLS protocol. However, the TLS protocol cannot fully guarantee that the content of the HTTP session is not tampered with. In the following, two related studies for improving the TLS protocol are explained.  

% \subsubsection{TLS-based Proof}

Ritzdorf et al., \cite{tlsn} propose TLS-N that is a TLS extension for providing a decentralized, seamless, and standardized internet-wide non-repudiation mechanism to securely share data feeds. TLS-N produces proofs about the content of a TLS session providing an efficient way of verification by third parties and blockchain-based smart contracts. Based on the generated proofs, TLS-N allows parts of a TLS session, e.g., passwords, to be hidden for increasing privacy while the remaining content becomes verifiable. There are requester, generator, and verifier, and the process starts with the establishment of a TLS connection and negotiation of the TLS-N parameters in the handshake. During the TLS session, a small TLS-N state holding information about the hash value of the previous records, an ordering vector (i.e., a bit vector encoding the interleaving requester and generator records), and a timestamp (to mitigate time-shifting attacks) are maintained and kept updated by the generator that signs its TLS-N state by its private key. Upon asking for the evidence by the requester, the evidence window that consists of records to be included closes, and the requester maintains full control of the included records in the proof, and by checking the proof the verifier can access the content of the TLS session. To reduce the size of proof, Merkle Tree is used, and if the session contains sensitive information to be hidden, independent random values called salts are needed that are derived from the TLS traffic secret using a record-based nonce. Hence, the verifier for the proof verification re-generates the evidence such that in the absence of sensitive information, it constructs the Merkle (and salt) Tree, otherwise, it produces the partial Merkle Tree with respect to the provided plain text, commitments, and hashes.

TLSNotary \cite{orc20} is a service that introduces third-party auditors for validating TLS session data exchanged between the client and server. It uses TLS protocol to help a client (auditee) with providing evidence of certain web traffic that occurred between the client and the server, to a third party (i.e., auditor). TLSNotary facilitates the verification of obtained data from external sources in an oracle against tampering through splitting the TLS master via the RSA encryption. Auditors hold a portion of the TLS secret key to generate the Message Authentication Code (MAC) key for setting up the HTTP session key intercepting the auditee's attempt to fabricate traffic from the server. Without the MAC key, the client cannot decrypt and authenticate the traffic received from the server. Upon commitment to the encrypted content of the server's response, the auditor provides the portion to the auditee to complete the authentication steps of TLS. Although promising, there are issues such as it is not supported by most web servers (websites must support TLS 1.0/1.1), and there has to be a trusted auditor for the process.

\begin{figure}[!b]
    \centering
    \includegraphics[keepaspectratio,scale=0.25]{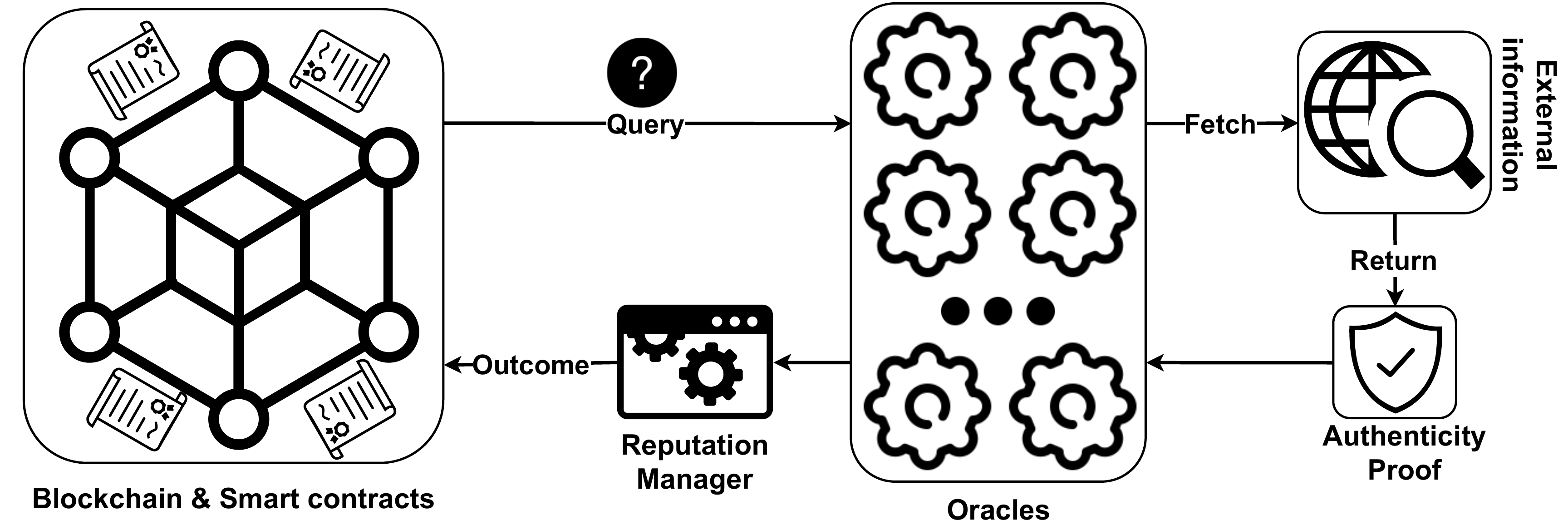}
    \caption{Reputation-based oracles overall structure. Malicious oracles can be blocked by the reputation manager, and authenticity proof may be attached to the data. }
    \label{fig:reputation_block}
\end{figure}

It should be noted that there are studies whose reputation-based oracle design does not practice authenticity mechanisms (e.g., \cite{orc10,orc28}) as the data sources are assumed to be trusted and the data integrity is intact during data retrieval. Figure \ref{fig:reputation_block} depicts the overall structure of reputation-based oracles assisted by the authenticity proof mechanisms, and oracles may return data to the blockchain without authenticity proof mechanisms.

%to ensure data given to the oracles are untampered. Authenticity proof is a mechanism to

% \subsubsection{Authenticity Proof}
% \label{sec:authproof}

The following sections present how these proofs are employed and developed for data authenticity in the reputation-based oracles.

\subsection{Software-based Proof}
\label{sec:sft_prf}
Guarnizo et al., \cite{pdfs} present PDFS as a data feed system that allows data to be authenticated over the blockchain without breaking TLS trust chains or modifying TLS stacks. Content providers can specify data formats to freely use, thus, data can be easily parsable and tailored for smart contracts. Also, PDFS provides content providers with a payment framework to be incentivized, but it does not allow content providers to misbehave by equivocating or censoring queries. They are defined as modification or deletion of the published content retrospectively, and influencing a contract execution by censoring some required content, respectively. In PDFS, content providers create authoritative contracts enabling other contracts to verify the authenticity of the content, and providing functionalities to mitigate misbehavior. Content providers then create a signed manifest containing information about the blockchain address, the authoritative interface, and metadata of the content. The manifest’s signature is computed with the help of the private key corresponding to the public key from the content provider’s TLS certificate allowing contract parties to verify the authenticity of the manifest directly. Content providers create a tamper-evident data structure (TDS) (i.e., Merkle Tree) storing served data entries including the manifest, and per each update the data structure is re-computed and its consistency proof is sent to the authoritative contract for validation. If contract parties wish to deploy relaying contracts (i.e., smart contracts require data feeds from external websites) they should find and agree on a content provider necessitating verification of its manifest and authoritative contract. The relying contract is called by a contract party, and interacts with the authoritative contract’s membership verification method for the produced data by the content provider. PDFS is resilient to attacks such as TLS Public Key Infrastructure (PKI) compromise and malicious content providers, as, in the former, data verification is done by the correct deployed authoritative contract while in the latter the TDS consistency is enforced by authoritative contracts noting that the content provider's response is visible.

Zhang et al., \cite{deco} present Decentralized Oracle (DECO) assisting users with proving data accessed via TLS that comes from a particular website, and provides statements (zero-knowledge proofs (ZKPs)) about the data. Authors argue that mandating installing TLS extensions at servers suffers from two issues, the first of which is that legacy compatibility becomes broken and reduces wider adaptability, and the next issue is the limitation of data exportability as the web servers determine what data to be exported resulting in censoring export attempts. Hence, the oracle does not require trusted hardware or server-side modification and provides a three-party handshake protocol to mitigate forging arbitrary TLS session data. This is due to the TLS nature generating symmetric encryption and authentication keys shared between users and web servers. Moreover, DECO reduces context-integrity attacks (i.e., specific data not only exists in the server's response but also it appears in the expected context) via a two-stage parsing scheme as attacks can be thwarted if the session content is structured and parsable. DECO consists of three phases; three-party handshake phase for the establishment of session keys in a special format for unforgeability, a query execution phase such that the server is queried for data based on a query built from the template with the private parameters, and a proof generation phase in which the query is proved that it is well-formed and the desired condition is satisfied by the response. In the first phase, the session key used in the TLS session with a server is distributed between the prover and verifier in a secret-share form. In the next phase, since the session key is secret-shared with prover and verifier, both are needed to interact with each other and execute a two-party computation protocol for the construction of TLS records encrypting the query. Finally, upon receiving a response from the server, the prover commits to the session by providing the ciphertext to the verifier to obtain the MAC to verify the response integrity and prove statements about it.

He et al., \cite{sdfs} present SDFS as a scalable data feed service employing a reputation-evaluation strategy for malicious node detection that leverages a blockchain to preserve the data processing. The proposed data feed service consists of smart contracts to invoke the interface for requesting the data, a server which has multiple nodes for data fetching, and an auditor blockchain assisted by TLSNotary \cite{orc20} to verify that obtained data is untampered. The service maintains a reputation mechanism for data feed nodes updated per data feed that maintains a verification pass and based on the pass, in each round the reputation value of the node is increased or decreased. 

\begin{table*}
\renewcommand\arraystretch{1.35}
\centering
\caption{The summary of software-based studies for reputation-based oracles}
\small
% \begin{adjustbox}{width=\linewidth}
\linespread{1}\selectfont
		\begin{tabular}{|p{1.5cm}<{\centering}|p{8.5cm}<{\centering}|p{2cm}<{\centering}|p{2cm}<{\centering}|p{1.8cm}<{\centering}|}
		\hline \bf{Literature} &
		\bf{Key Research Outcomes} &
		\bf{Limitation} &
		\bf{Authenticity} &
		\bf{Confidentiality}\\
		\hline
		PDFS \cite{pdfs} & A data feed system for data authentication over the blockchain without breaking TLS trust chains or modifying TLS stacks.& None & \checkmark& \xmark \\ \hline
        Ritzdorf et al., \cite{tlsn}& Providing a TLS extension that is compatible with TLS 1.3 and produces proofs for the content of a TLS session.&Adding minor overhead to the TLS&\checkmark&\checkmark\\ \hline
        TLSNotary \cite{orc20} & Verification of the obtained data from external resources against tampering through splitting the TLS master.  & Only works with TLS 1.0/1.1.& \checkmark& \xmark \\\hline
        Zhang et al., \cite{deco} & Requires no-server side cooperation, and having support for TLS 1.2 and 1.3. & None & \checkmark & Partially\\\hline
        He et al., \cite{sdfs} &Reputation-based scalable data feed service assisted by TLSNotary for data verification against tampering. & Restricted to TLSNotary limitations& \checkmark &\xmark\\\hline
        Bridge Oracle \cite{bridgeoracle} & A Dedicated oracle technology on the Tron network providing the ability to attach authenticity proofs to the requested data.& Unreliable \cite{bridgescam} & \checkmark &\xmark\\\hline
        JustLink \cite{justlink} & A decentralized oracle where a single result is calculated by an aggregator contract. & Not resilient to data tampering and usage of defined trusted sources & \xmark &\xmark\\\hline
        
        Pyth Network \cite{pythnetwork} & A Solana-based cross-chain market that consists of different entities such as delegators for providing a confidence level in the data from a particular Pyth data provider. & Not resilient to data tampering & Partially &\xmark\\	\hline
        API3 \cite{api3} & A decentralized autonomous organization aiming for creation and monetizing a decentralized API network to act as a bridge for connecting blockchains to the existing data provider APIs governed by API3 token. & Not resilient to data tampering & Partially &\xmark\\	\hline
        PolkaOracle \cite{polkaoracle} & A Polkadot-based oracle that employs POT for governance as well as Substrate 2.0 Off-chain Worker for securely integration of data to the blockchain applications. There is the data source layer that uses techniques (e.g., filtering or screening) for accuracy and authenticity of the data. & Not resilient to data tampering & \checkmark &Partially\\	\hline
    \end{tabular}
    % \end{adjustbox}
    \label{tbl:rep_soft}
\end{table*}

Bridge Oracle (BRG) \cite{bridgeoracle} is an \emph{unsuccessful} public oracle technology on the Tron network \cite{tronnetwork}. The Bridge oracle provides the ability to attach authenticity proofs, and deals with a variety of APIs and parsing helpers while employing TRON (i.e., the Tron network token) and project-purpose tokens on the network for payments. The bridge oracle consists of three main smart contracts; (1) the Bridge API contract (e.g., public, decentralized, and enterprise) for connecting the client smart contract to the Bridge oracle, (2) The bridge oracle address resolver that is in charge of redirecting requests to the correct services (e.g., public oracle system or decentralized oracle system), and (3) the bridge oracle connector for processing requests and outputting specific data to be accessed by oracle data carriers. This oracle can deal with three types of time-variant requests such as one time, scheduled, and open ending queries, and for open ending queries three technologies such as web socket protocol, long polling method, and recursive HTTP(s) request method are proposed. A queuing system is employed for the off-chain bridge oracle architecture to balance the load of data carriers, a random access memory (RAM)-based database is used for temporarily logging queries to avoid losing queries during load condition while the query's feedback is stored in the permanent database.

JustLink \cite{justlink} is a decentralized oracle network that is an under-development project deployed on the Tron network. Data requests interact with the on-chain JustLink open-source and verifiable interface that includes smart contracts. There is an on-chain aggregator contract for which users choose nodes and services. The final result is computed and finalized through \emph{trusted} sources for requesting contracts. Based on the requirements, oracles are selected and the aggregator that can be different for each demand, outputs the result by, e.g., calculating the weighted average after removing abnormal data. Off-chain data are obtained by nodes separately, and a single result is finally calculated in the aggregator contract. Each assignment can be divided into subtasks (e.g., HTTP requests or JSON parsing) that are customizable within external adaptors as services with a minimal REST API, and subtasks pass their results to the next subtasks as they run end to end to obtain the final result. In JustLink, freeloading issue is handled by commit/reveal, and JustLink employs the token named Just (JST) for paying JustLink Node operators that retrieve data from off-chain data feeds. In addition, JustLink plans to devise a reputation system for controlling the quality of oracles and leverages a certification service to mitigate Sybil attacks in which statics from the validation system are collected. Also, the service performs after spot-checking of on-chain answers.

Pyth network \cite{pythnetwork} is a solution for providing a cross-chain market of verifiable data in a decentralized way that is powered by Solana-based blockchain \cite{solana}. Heavy and fast processing of the data can be achieved by Solana as it is recognized to be the only chain providing the computing bandwidth. Pyth network employs formal likelihood methods to output a suitable price considering all the received information through high-quality data providers. Also, the Pyth network is capable of including information such as historical quality or potential stake at risk. The Pyth network consists of entities, each of which is responsible for a particular task; data providers who can be data owners or source for providing new datasets and obtaining data on-chain, respectively. Delagotors as the other entity is in charge of providing a confidence level in the data from a particular Pyth data provider which can be done through evaluation and consideration of their historical performance and accuracy, and can be rewarded or penalized. Also, there are curators for being in charge of determining which data should be sourced through paying tokens into a bonding curve to signal interest. Increasing the interest in symbols will lead to receiving the greatest share of rewards which incentivizes the data providers for providing prices for them.

API3 \cite{api3} is recognized as a decentralized autonomous organization (DAO) that aims to create and monetize a decentralized API (dAPI) network to act as a bridge for connecting blockchains to the existing data provider APIs. Each dAPI has oracles managed by decentralized API providers and holds API3 token enabling holders through staking the tokens to practice governing rights over the API3 DAO with the rewarding opportunity. Hence, unlike the general oracles, API3 aims to be governed by DAO that means the API3 ecosystem players are in charge of securing the network. As first-party oracles build a data feed, over-redundant decentralization would not be necessary and it would be more immune to attacks resulting in better transparency. Also, a user could easily identify data owners as on-chain identities of API providers are published via off-chain channels, and funds are transferred to them because of performing the actual work instead of fee-paying third-party oracles. Data is signed by the API provider and becomes accessible via a regular API endpoint that third-party oracles can query to fetch the data, and the authenticity of the data is verified by their public keys.

PolkaOracle \cite{polkaoracle} is a Polkadot-based oracle network that aims to be a community-driven oracle system, and employs POT as the native token in the network to pay the data providers, to govern the network via voting, and deposit purposes. In comparison to Chainlink \cite{chainlink2}, PolkaOracle provides flexibility and reliability through leveraging Substrate 2.0 Off-chain Workers for the infrastructure that can act as parachain or parathread for connecting to the Polkadot blockchain. It uses on-chain operations such as on-chain computation, encryption and decryption, data verification, and random challenge for credible and reliable real-time data feeding. PolkaOracle has a layered architecture; (1) cross-chain application layer for providing data interfaces based on the cross-chain technology for applications and public tools (e.g., data display panels), (2) the on-chain infrastructure for security and transparency of the network. Also, (3) there are Off-chain Workers to securely integrate data to the blockchain applications as it utilizes verifiable random functions (VRF) to randomly select network nodes for off-chain calculation and verification for making sure the data is not wrong or tampered. Finally, there is a (4) data source layer that is responsible for obtaining third-party off-chain data via APIs, and employs techniques (e.g., filtering or screening) to ensure accuracy and authenticity of the data.

Table \ref{tbl:rep_soft} presents the summary of studies that employed software-based strategy to provide proof in the reputation-based oracles. A majority of studies may fail to satisfy the data integrity as data could be tampered during transmission between source and oracles.

\subsection{Hardware-based Proof}
\label{sec:hdw_prf}

Schaad et al., \cite{orc29} present a Hyperledger-based blockchain design with a local secure element (Wibu CmDongle for cryptographic software protection \cite{wibudongle}) as an external hardware-based oracle. They applied a use-case study where a 3D printer is rented and loaded with a specified amount of printing credit. Per each print, the local counter embedded in the dongle is decreased and in parallel, the counter unit is maintained on the Hyperledger blockchain. Upon meeting the threshold a chaincode on the blockchain is triggered to set the counter again based on the payment processing, and update the local counter on the device for further printing.

Ledger proof \cite{ledgerproof} leverages hardware wallets owned by the Ledger company. These hardware devices employ Blockchain Open Ledger Operating System (BOLOS) that provides Software Development Kit (SDK) and enables developers to code applications (i.e., cryptocurrency wallets) to be installed on the hardware. It provides an isolated environment as each application has its memory region operating in the user mode and interacting with the operating system in the superuser mode.

Town Crier \cite{orc9} is an authenticated data feed system that acts as a bridge between smart contracts on the Ethereum blockchain and commonly trusted websites' data known as datagrams. Moreover, Town Crier ensures \emph{confidentiality} that is referred to as requesting private data with encrypted parameters, e.g., accessing online accounts. Town Crier employs a combination of a front-end smart contract and Intel Software Guard Extension (SGX) technology; a set of instructions granting hardware protections on the user-level code. Town Crier has three components; the town crier contract, the enclave, and the relay. The enclave and relay reside on the Town Crier server while the contract is on the blockchain. The relay functionality is defined as handling the network traffic on behalf of the enclave. The front-end smart contracts respond to requests from contracts on the blockchain with the attestation holding characteristics of datagram parameters, HTTPS website, and the time frame. A relaying contract can verify the datagram considering trust in the SGX security, Town Crier code, and validity of the source data in the time frame. In addition to data authenticity, Town Crier proves gas sustainability (i.e., Ethereum service never runs out of gas), and trusted computing base code minimization by authenticating the enclave outputs on the blockchain. There are still some issues; (1) an enclave requires network capability (can be done by splitting TLS code between the enclave and untrusted host environment), and  (2) compromising a single website or an enclave that is addressed by the majority voting.  

Android proof \cite{androidproof} uses Google technologies such as SafetyNet Software Attestation and Android Hardware Attestation (implemented in a Trusted Execution Environment (TEE)\footnote{It is defined as a computational environment which is heavily isolated from the main operating system running on a device.}) to notarize web pages or certify data served by HTTPS APIs. The former evaluates the application runs on a safe and not rooted physical device, i.e., unmodified root certificate authorities (CA). Besides, it checks the application code hash and makes sure that the source is untampered. The latter verifies that the device is running on the latest OS version for preventing any potential exploits. Hence, both technologies assure that the device is a secure environment for making an untampered HTTPS connection with a remote data source. When a request of Android proof becomes available, the given URL by the user is forwarded to the Android device, an HTTPS connection is established, and the entire HTTP response is retrieved. The SHA256 hash of the response is signed with the hardware attested key pair available on the device. SafetyNet's API is called by the service application, and the nonce parameter of the API becomes SHA256 hash of the HTTP response key, the signature, and the request identifier formatted as a JSON Web Signature (JWS). By full validation of the proof, the data in the HTTP response is parsed and distributed to the user with the SafetyNet Attestation Response and Hardware Attestation Object. Although promising, there is a quota (10k requests per day) for Google SafetyNet API that limits the scalability of the system.

Woo et al., \cite{orc30} propose a distributed oracle for safely importing time-variant data into the blockchain where the response time is important. The proposed oracle employs multiple oracles to support data availability as well as data integrity with the help of Intel SGX. Each oracle verifies data pulling procedure such that other oracle nodes pull data from external data sources through remote attestation\footnote{It is a method by which hardware and software configuration of devices are authenticated to remote hosts.} provided by Intel SGX. The proposed oracle resolves malicious oracle with the help of a reputation system to block selfish oracles from obtaining benefits.

Edenchain \cite{orc17} is a permissioned blockchain platform technology for capitalizing assets of any form into a token. Edenchain employs namespaces with Merkle Tree and isolates transactions based on the namespaces (the type of transactions) for increasing the performance. Also, it employs Proof-of-Elapsed-Time (PoET) implemented in the SGX enclave as a consensus algorithm using CPU commands to randomly select a leader with the smallest wait time without requiring to consume excess energy for solving a hash problem. Edenchain has three layers; a distributed ledger layer based on the Hyperledger for storing data, a validation layer for execution and verification of a transaction, and a bridge layer for securely importing required data by on-chain smart contracts. In the bridge layer, on-chain and off-chain nodes exist, and a reliable communication between nodes is managed by E-Protocol that implements an encryption technique called Elliptic Curve Cryptography–Threshold Cryptography (ECC-TC). Threshold cryptography is a protocol with a cooperative property in which data for decryption is shared among participants. On-chain nodes interact with the smart contract while off-chain nodes are designed for interaction with the external system and making a connection between the on-chain and off-chain modules. The Edenchain uses the E-Bridge layer for fetching data from multiple data sources, and encrypts this data, and leverages the median voter theorem (MVT) to secure trust and improve security. The core technology of the bridge layer is E-Bridge that has a modular design with components such as an oracle server and a SGX enclave located off-chain for serving requests and providing the trusted execution environment. The other components are an executor located on chain for transaction execution, and E-Oracle for forwarding data access requests from the smart contracts. The E-Oracle has a client and server such that the former provides parameters to be executed on the server, and the server executes external data requests on separate spaces named SGX enclaves for security enhancement. It also collects external data, selects appropriate values, and forwards the values to the client. The E-Oracle servers can have discrete type data and continuous type data, each of which is evaluated by the majority voting, i.e., the most common value and median voter theorem (MVT). In MVT, the result is chosen by a median voter and consensus algorithms that make MVT suitable for continuous data types. 

Hearn presents Corda \cite{corda} as a decentralized global database platform without a mining concept for recording and processing financial agreements. Corda aims to provide a distributed ledger consisted of mutually distrusting nodes to let a single global database keep the deal states and obligations between people and institutions free of disparate ledgers synchronization. The Corda employs SGX enclaves for attestation and supports smart contracts and leverages cryptographic hashes for data and parties identification. It also defines state objects as a digital document recording the existence, content, and the current state of an agreement between two or more parties. It employs a consensus algorithm for transaction validity (the parties who are involved in) by independently checking the associated code runs successfully that has the required signatures and refereeing transactions are valid and unique. Contracts are executed in the Java virtual machine which eases reusing the existing code in the contracts. The Corda network has one or more notary services, and zero or more oracle services. Oracles are implemented in two ways; by using commands in which a fact is encoded in a command embedded in the transaction itself, and the oracle becomes a co-signer of the entire transaction, and if a transaction includes the fact, it must be returned to the oracle for signing. The other way is using attachments that facts are encoded as the attachment and are considered separate objects to the transaction and are referred by the hash. The transaction content becomes accessible from oracles by employing the Merkle Tree that reveals only necessary parts of transactions.

Provable \cite{orc8} is a platform-agnostic bridge between the blockchain and the internet and behaves as a data carrier to provide a reliable connection between Web APIs and dapps. Provable employs cryptographic proofs such as TLSNotary to enforce reliability, and the platform can be used in public and private blockchains, and even in non-blockchain contexts. Provable also provides the ability for users for encrypted queries (does not support private or custom datagrams) via the provable public key, and the plain-text queries can be protected by the Elliptic Curve Integrated Encryption Scheme. Moreover, Provable presents ProofSheild assisting smart contracts for verification of on-chain authenticity proofs provided by Provable. It provides tools and services for connecting oracles (data providers) with distributed applications, however, it is more suitable for centralized data centers as oracle solutions. Provable facilitates the data verification through returning data with a document named authenticity proof which can be produced by technologies such as auditable virtual machines and trusted execution environments. Provable assumes that data fetched from the sources are genuine and untampered, and provides a variety of parsing helpers to extract a value of a data type. Provable employs Android proof for authenticity proof, and the verification and proof process consists of a series of verification such as SafetyNet Authenticity verification, SafetyNet Response verification, and Hardware Attestation verification. Additionally, these services ultimately rely on the reputations of their (small) providers to ensure data authenticity. 

Gray et al., \cite{bletchley} present Bletchley; a Microsoft Azure-based enterprise consortium blockchain ecosystem that is similar to the Microsoft Coco framework. Russinovich et al., \cite{cocoframework} present the Coco framework as an open-source system that is a high-scale and confidential blockchain specifically designed for the confidential consortium, and employs Intel SGX and Windows Virtual Secure Mode (VSM) to create a trusted execution environment. Bletchley consists of two major components; (1) blockchain middleware providing core service functionalities in the cloud such as operation management and data services and (2) Cryptlets that enables secure communication between Microsoft Azure, middleware, and customers for providing information for the transaction execution. Cryptlets are assumed as a secure blockchain middleware tier that provides the oracle functionality, and are defined as off-chain components written in any language. They execute within a secure and trusted container, and communicate with using a secure channel. They can be used in smart contracts by an adaptor (CryptoDelegate) such that the adaptor in the smart contract calls Cryptlets which extends the secure and authentic envelope for the transaction. Two types of Cryptlets exist; utility and contract, the former is organized into services or libraries to provide common functionalities, e.g., encryption. In contrast, the contract Cryplets run within an enclave and provide all the execution logic securely storing data in the smart contract and can function as autonomous agents or bots for interaction with the off-chain world (i.e., acting as multiple oracles) but maintaining the integrity of the blockchain and smart contracts. Cryptlets can be accessed by a trusted attested host and can employ enclaves for process isolation and encryption. Cryptlets can be event-driven or control-driven as the former provides notifications based on events and securely passes data. The latter is followed by Cryptlet Contracts allowing Cryptlets to perform the required business logic.

\begin{table*}
\renewcommand\arraystretch{1.35}
\centering
\caption{The summary of hardware-based studies for reputation-based oracles}
\small
% \begin{adjustbox}{width=\linewidth}
\linespread{1}\selectfont
		\begin{tabular}{|p{1.5cm}<{\centering}|p{8.5cm}<{\centering}|p{2cm}<{\centering}|p{1.5cm}<{\centering}|p{2cm}<{\centering}|}
		\hline \bf{Literature} &
		\bf{Key Research Outcomes} &
		\bf{Limitation} &
		\bf{Authenticity} &
		\bf{Confidentiality}\\
		\hline
		Schaad et al., \cite{orc29}& A Hyperledger-based blockchain design with a local secure element (Wibu CmDongle). &Hardware requirement&Partially&\xmark\\\hline
		Town Crier \cite{orc9} & An enclave-based oracle for pull-based data provisioning assisted by Intel SGX technology.  & Hardware (Intel CPUs) dependent as a trusted third party & \checkmark & \checkmark\\\hline
		Ledger proof \cite{ledgerproof} & Providing functionalities to the developers for coding cyrptocurrency wallets. & Hardware requirement and prone to fail \cite{ledgerfail} & \xmark & Partially \\\hline
		Android proof \cite{androidproof}& Notarizing web pages or certifying data that is served by HTTPS. & Operating system based & \checkmark & Partially \\\hline
		Woo et al., \cite{orc30}& Presenting a distributed oracle for time-variant data to be recorded onto the blockchain by using enclaves (Intel SGX). & Hardware (Intel CPUs) dependent& \checkmark& Partially\\\hline
		Edenchain \cite{orc17}&A permissioned blockchain platform technology for capitalizing assets of any form into a token. &Hardware (Intel CPUs) dependent&\checkmark&Partially\\\hline
		Corda \cite{corda}&A decentralized global database platform for recording and processing financial agreements. &Hardware (Intel CPUs) dependent&\checkmark&Partially\\\hline
		
		Provable \cite{orc8} &A platform-agnostic bridge between the blockchain and the internet, i.e., web APIs and dapps, that employs Android proof for data authenticity. &Centralized and operating system based&\checkmark&Partially \\\hline 
		
		Bletchley \cite{bletchley} &A Microsoft Azure-based enterprise blockchain that uses Cryptlets to provide oracle functionality defined as off-chain components written in any language, and execute within a secure and trusted  container, and communicate with using a secure channel.&Hardware (Intel CPUs) dependent&\checkmark&Partially \\\hline
		
		Chainlink \cite{orc5}& A general-purpose and token-based framework for building secure decentralized input and output oracles for complex smart contracts on any blockchain.&Hardware (Intel CPUs) dependent&\checkmark&\checkmark\\	\hline
		
% 		Ampleforth \cite{ampleforth}& An Ethereum-based blockchain for incentivizing a network of users to maintain a value of crypto-asset equal to the U.S. dollar. &Hardware (Intel CPUs) dependent&\checkmark&\checkmark\\	\hline
    \end{tabular}
    % \end{adjustbox}
    \label{tbl:rep_hard}
\end{table*}

%%old place for rep table

Chainlink \cite{orc5} proposes a general-purpose and token-based framework for building secure decentralized input and output oracles for complex smart contracts on any blockchain. A Chainlink node can have multiple external adapters for different data sources, and its token protocol is blockchain agnostic that can run on different blockchains simultaneously. The Chainlink has two major components that are on-chain and off-chain components. The on-chain component has contracts such as (1) reputation; for tracking the performance metric, (2) order-matching; taking and logging a proposed service level agreement, and collecting bids from oracle providers. Also, the last contract is (3) aggregating that is in charge of response collection from oracle providers to calculate the final collective result of the query and is also responsible for feeding the oracle provider metrics, i.e., the reputation contract. The Chainlink on-chain component follows a workflow that is defined as query parameters, number of needed oracles, and reputation and aggregating contracts for Service Level Agreement (SLA) proposal are prepared by an oracle service purchaser. Then, the purchaser submits the SLA to an order-matching contract on which oracle providers based on their capabilities and service objectives filter the SLAs. The Chainlink nodes decide whether to bid (i.e., stake) on the proposal or not, and only bids from nodes satisfying the SLA’s requirements are accepted. The bid on a contract means commitments within a bidding window and is subject to penalties because of misbehavior. Once the biding window ends and enough qualified bids are received, the requested number of oracles is selected from the pool of bids, and the finalized SLA record is created and selected oracles are notified for performing the assignment and reporting. The aggregating contracts calculate a weighted answer, and the validity of each oracle response is reported to the reputation oracle. For the off chain contract, Chainlink has components such as core, external adaptors, and subtask schemes. The core node software is responsible for interacting with the blockchain or work (i.e., assignments) balancement across multiple external services. Each assignment consists of subtasks that is processed as a pipeline. Also, custom subtasks can be created by adaptors defined as external services with a minimal REST API. Chainlink has a validation system for monitoring the on-chain oracle behavior in terms of availability and correctness, and provides performance metrics. It also has a reputation system for collecting user ratings of oracle providers and nodes and presenting their historical performance. It also has a certification service that is responsible for endorsements of high-quality oracle providers and employs an optional contract-upgrade service to create a new set of oracle contracts in case of vulnerabilities.

The Chainlink technology has also progressed \cite{chainlink2} toward providing key oracle functions, e.g., an extensive collection of on-chain financial market data or verifiable randomness backed by on-chain cryptographic proofs. %, supplying on-chain data feeds for performing on-demand audits of tokenized asset reserves, and external adaptors for connections to any off-chain resource or API. 
Chainlink 2.0 is a decentralized oracle network (DON) for the creation of a decentralized meta layer for enhancing smart contracts. There are hybrid smart contracts where DONs offer capabilities to fill in the blockchain limitations, and and being connected to the off-chain systems. By the advanced off-chain computation, DONs provide a blockchain-agnostic gateway for smart contracts not only for the off-chain access but also providing an execution code environment to address blockchain limitations. The use of Chainlink can also be seen in Ampleforth \cite{ampleforth} that is recognized as a piece of software running on the Ethereum blockchain aiming to incentivize a network of users to maintain a value of crypto-asset equal to the U.S. dollar. The Ampleforth employs a token called AMPL such that its supply is adjusted programmatically by the software that reduces the reliance on deposits or issuing and redeeming debt. This process is called ``rebasing'' and takes place every 24 hours in a way that if the demand for AMPL tokens is high, and the price of each AMPL token exceeds \$1, there will be an increase in supply. Otherwise, the supply will be adjusted and will decrease, hence, the AMPL token is recognized as a cryptocurrency token that is elastic and non-dilutive. In other words, despite changes in the supply, users keep possession of the same proportion of the overall supply. %, and it adjusts the supply of AMPL managed by the software daily to maintain price parity with the U.S. dollar. 

Table \ref{tbl:rep_hard} illustrates the usage of hardware for data authenticity proof may lead to limiting the oracle to employ specific hardware. Although promising, it is not a generalized strategy and does not provide flexibility.

\subsection{Proofless}
Wang et al., \cite{orc10} present an oracle based on Application Specific Knowledge Engines (ASKE) that is a framework to acquire and analyze information. Open-source information in specific domains is collected and unified, and the framework analyzes the collected data in different dimensions by the integration of data analysis methods. The analysis uses knowledge configuration files (KCF) for specifying keywords, search sequences, topics, and schedules for query processing, and helping users with accurately finding the required information. This framework facilitates data collection from authoritative websites via web crawlers automatically and aggregates data off-chain to produce the final results for sending to the smart contracts. In this proposed oracle, authoritative domains for fetching the information are recognized by asking domain experts (i.e., evaluation of their reputations), then the oracle determines the domain of a request, and with the help of the ASKE framework, the results are extracted, analyzed, and returned to the blockchain. The proposed oracle is not decentralized, and it does not practice data authenticity mechanisms, and the oracle lacks heterogeneous data sources for intelligent data collection.

Al Breiki et al., \cite{orc28} present a decentralized access control for IoT data which is assisted by blockchain and trusted oracles. It leverages smart contracts to shift access management toward a decentralized, secure, and scalable management for IoT data access. It employs multiple oracles to provide decentralized but trusted source feeds for IoT data. The proposed system consists of entities; admins for user control access, end-user (i.e., dapps or wallet), smart contracts for verifying IoT user data access, oracles for providing information about the registered oracles, aggregator, and the reputation smart contract. They are responsible to send a data request to the set of oracles, and compute the hashes for the requested data. They then compare hashes and report to the reputation smart contract for averaging.

\begin{table*}
\renewcommand\arraystretch{1.35}
\centering
\caption{The summary of proofless-based studies for reputation-based oracles}
\small
% \begin{adjustbox}{width=\linewidth}
\linespread{1}\selectfont
		\begin{tabular}{|p{1.5cm}<{\centering}|p{8.5cm}<{\centering}|p{2cm}<{\centering}|p{1.5cm}<{\centering}|p{2cm}<{\centering}|}
		\hline \bf{Literature} &
		\bf{Key Research Outcomes} &
		\bf{Limitation} &
		\bf{Authenticity} &
		\bf{Confidentiality}\\
		\hline
		 Wang et al., \cite{orc10} &  Leveraging Application Specific Knowledge Engines (ASKE) for information acquisition and analysis. &Not decentralized and lack of data authenticity mechanisms & \xmark&\xmark \\\hline
		Al Breiki et al., \cite{orc28} &A decentralized access control for IoT data that is assisted by blockchain and trusted oracles. &No authenticity proof mechanism and relying on truthful oracles.&Partially& \xmark\\\hline
		Fujihara \cite{orc27} &Presenting open data platform that is assisted with a decentralized oracle for correct information extraction.& No authenticity proof mechanism& Partially &\xmark\\\hline
		Pedro et al., \cite{witnet} &A decentralized oracle network that runs on a blockchain with a native protocol token. It employs truth-by-consensus protocol for obtaining the truth, and uses miners for retrieving, attesting, and delivering (RAD) of web contents which heavily depends on the reputation. &No authenticity proof mechanism&\checkmark&Partially\\	\hline
    \end{tabular}
    % \end{adjustbox}
    \label{tbl:rep_proofless}
\end{table*}

Fujihara \cite{orc27} presents a blockchain-based open data platform that relies on the mobile crowdsourcing. It employs a decentralized oracle to extract the true binary information for saving recorded onto the blockchain via an algorithm called Expectation-Maximization (EM) algorithm. In the proposed platform, there are task requester, workers, and tasks, and the algorithm is done in $E$ and $M$ steps such that in the former step, for each task, the corresponding ${E_i}$ value is determined by the Bayes' theorem. In the $M$ step, the probability of the correct answer for tasks considering the E-step is calculated. By the step repetition, the estimated values gradually converge to fixed values resulting in the determination of workers' reliability score. This score is used for incentivizing workers that is proportional with respect to the score.

Pedro et al., \cite{witnet} present Witnet which is a decentralized oracle network that runs on a blockchain with a native protocol token for incentivizing miners. It enables any software to retrieve content published at any web address (HTTP/HTTPS) with a complete and verifiable proof of its integrity while being immune to the Sybil attacks and laziness (e.g., verifier's dilemma). The ledger on the Witnet is based on directed acyclic graph (DAG) where multiple blocks can exist at a time while enforcing a legit ledger. In addition, miners that are referred to as witnesses earn by retrieving, attesting, and delivering (RAD) web contents to the users via a deterministic algorithm that heavily relies on the reputation. This reputation is affected by demurrage gradually meaning a deduction on reputation scores is proportionally applied. The Witnet follows a strategy to incentivize network participants to take part in the RAD requests as their reputation is lost if they start hoarding their points. Hence, the Witnet encourages participants to become witnesses and participate in the outcome of RAD requests. They are required to work honestly as contradicting with the majority of miners would lead to losing the reputation. In the Witnet, witnesses compete with each other to earn a reward, and with respect to their mining power, rewards become proportional to the previous honesty and trustworthiness, i.e., their reputations. The Witnet employs truth-by-consensus protocol to obtain the ``agreed truth''. This protocol is based on singular value decomposition (SVD) to analyze a matrix that contains all the claims produced during epochs. Moreover, the network scalability is guaranteed by the sharding feature of the Witnet in addition to allowing clients to choose a number of witnesses for the RAD tasks. If the Witnet is coupled with a decentralized data storage (DSN), a digital knowledge ark can be built that is immutable and resistant to censorship. Blocks are created periodically and it does not depend on the time spent on solving the proof of work challenge by the fastest miner. Miners use a scriptable headless browser that has no interface to retrieve information from websites. In addition to clients and miners, there are bridge nodes that are in charge of watching other blockchains in case of RAD requests and replicating the results upon requests.

Table \ref{tbl:rep_proofless} presents the summary of proofless strategies, and it illustrates that they barely utilize authenticity mechanisms for verifying data integrity.

\section{Future Research Directions}
\label{sec:frd}
This section presents future research directions in the blockchain oracle design and usage. Although it may sound blockchain oracles are well-studied, there are still unaddressed research and technical questions in practice.

\subsection{Operating Cost \& Speed}
Smart contracts use resources for execution, e.g., gas in the Ethereum blockchain, hence, it necessitates to develop not only very efficient and optimal code for smart contracts but also provides a faster response time for incoming queries. While there have been studies for developing cost-effective blockchain-based applications (e.g., \cite{gascost}), there still a need for designing high-performance blockchain oracles. For example, Chainlink has recently presented Off-Chain Reporting (OCR) that significantly improves data computation across Chainlink oracles while reducing the operating cost \cite{chainlinkocr}. Moreover, the design and deployment of oracles should be relied on employing high-performance and low transaction fee blockchains, e.g., Solana or Polkadot blockchains.  

\subsection{Decentralized Oracles \& Security}

Although presented oracle techniques may sound sophisticated and novel, they still require data integrity and authenticity mechanisms for enforcing security and privacy. Although decentralized oracles provide benefits for data acquisition, data security challenges are still the issues. Hence, the design of oracles should provide an acceptable level of data integrity, security, and privacy. In Section \ref{sec:voting_oracle}, we discussed different strategies for the voting-based oracles, however, the majority, if not all of them, barely employed authenticity proof mechanisms for data integrity and correctness.

Moreover, with the ever increasing introduction of blockchain oracles and their customized tokens to the market for investment, detecting a legitimate project is of great importance and requires more attentions from the research prospective. For example, Mate Tokay co-founder of bitcoin.com has filed legal action against the Bridge.link founders due to misleading Bridge token (BRG) investors \cite{bridgescam}. Hence, such issues should be further investigated and a set of requirements for identifying a legitimate blockchain oracle should be defined.    

\subsection{Blockchain-agnostic Oracles}
Majority of studies presented in Sections \ref{sec:voting_oracle} and \ref{sec:reputaion} employed a certain blockchain (i.e., Ethereum blockchain). In contrast, there are few studies that have presented blockchain-agnostic oracles, e.g., Band protocol \cite{bandchain} or Chainlink \cite{chainlinkocr} oracle. Therefore, blockchain oracles adaptability should be studied further to propose a set of crucial rules and requirements for developers. Also, it will make the blockchain oracle design more flexible and functional to the constant changes in such a growing ecosystem assisting decentralized applications with execution across multiple blockchains. This will benefit users to easily handle transactions without the need for using different exchange platforms. 

\subsection{Query Types}
Oracles designed for fetching off-chain information should be capable of dealing with different query types, e.g., binary, scalar, or categorical. In fact, processing non-binary query types is challenging as the diversity of responses may be big enough, hence, it requires techniques to efficiently manage data aggregation at the blockchain side. Moreover, there should be reliable while fast authenticity proof mechanisms to be attached to the data for data integrity verification.

% \cite{cocoframework}

\section{Conclusion}
\label{sec:conclusion}
Blockchain technology has disrupted digital interaction in our economy and society in the last few years. Blockchain as a form of distributed ledger technology has enabled data to be shared among nodes connected over the internet. In addition, by the introduction of smart contracts to the blockchain, programmability is added to this disruptive technology and has changed the software ecosystem by removing third parties for administration of (non)business purposes. Although promising, blockchain and smart contracts do not have access to the external world, hence, they need \emph{trusted} services referred to as blockchain oracles for sending and verifying external information to smart contracts. This paper presented an overview of blockchain oracles by categorizing them based on the feedback (i.e., the blockchain oracle outcome) into two major groups: voting-based oracles and reputation-based oracles. While the first group leverages voting strategies, e.g., stake on outcomes, for data aggregation and outcome, the latter considers reputation, e.g., the powerful participant(s) on the network, for choosing the oracle for reporting the outcome to the requester. Oracles may employ authenticity proof mechanisms to check the correctness and integrity of the obtained data from external sources. Our discussion and review showed that the existing strategies should keep the integrity of data obtained from external resources, and oracles honestly work toward providing the truth back to the blockchain and smart contracts. Moreover, although existing studies sound promising, blockchain oracles are still in need of further research from different perspectives such as operating cost, processing speed, security, and handling different query types.

\bibliographystyle{IEEEtran}
\bibliography{main}
\end{document}